\documentclass[11pt,preprint]{aastex}
\newcommand{\etal}{{\it et~al.}}

\usepackage{rotating}
\usepackage{longtable}

\begin{document}

\title{Asteroid family identification using the Hierarchical Clustering Method and WISE/NEOWISE physical properties}

\author{Joseph R. Masiero\altaffilmark{1}, A. K. Mainzer\altaffilmark{1}, J. M. Bauer\altaffilmark{1,2}, T. Grav\altaffilmark{3}, C. R. Nugent\altaffilmark{4}, R. Stevenson\altaffilmark{1}}

\altaffiltext{1}{Jet Propulsion Laboratory/Caltech, 4800 Oak Grove Dr., MS 183-601, Pasadena, CA 91109, {\it Joseph.Masiero@jpl.nasa.gov; amainzer@jpl.nasa.gov; James.Bauer@jpl.nasa.gov; Rachel.Stevenson@jpl.nasa.gov}}
\altaffiltext{2}{Infrared Processing and Analysis Center, Caltech, Pasadena, CA}
\altaffiltext{3}{Planetary Science Institute, Tucson, AZ {\it tgrav@psi.edu}}
\altaffiltext{4}{Department of Earth and Space Sciences, University of California, Los Angeles, CA {\it cnugent@ucla.edu}}

\begin{abstract}
Using albedos from WISE/NEOWISE to separate distinct albedo groups
within the Main Belt asteroids, we apply the Hierarchical Clustering
Method to these subpopulations and identify dynamically associated
clusters of asteroids.  While this survey is limited to the $\sim35\%$
of known Main Belt asteroids that were detected by NEOWISE, we present
the families linked from these objects as higher confidence
associations than can be obtained from dynamical linking alone.  We
find that over one-third of the observed population of the Main Belt
is represented in the high-confidence cores of dynamical families.
The albedo distribution of family members differs significantly from
the albedo distribution of background objects in the same region of
the Main Belt, however interpretation of this effect is complicated by
the incomplete identification of lower-confidence family members.  In
total we link $38298$ asteroids into $76$ distinct families.  This
work represents a critical step necessary to debias the albedo and
size distributions of asteroids in the Main Belt and understand the
formation and history of small bodies in our Solar system.
\end{abstract}

\section{Introduction}
In the first publication in this series \citep[][hereafter:
  Mas11]{masiero11} we presented the preliminary results for Main Belt
asteroids (MBAs) from the Wide-field Infrared Survey Explorer (WISE)
thermal infrared all-sky survey \citep{wright10} and the NEOWISE Solar
system enhancement to the core WISE mission \citep{mainzer11nw}.
Mas11 also presented the observed albedo and diameter distributions
for asteroid families drawn from the overlap between the set of
objects detected by NEOWISE and the families identified by
\citet{nesvornyPDS} using the Hierarchical Clustering Method
\citep[HCM,][]{zappala90,zappala94,zappala95,bendjoyaAIII}.  In this
paper we perform new analysis of the Main Belt using HCM, taking into
account dynamical associations as well as asteroid albedo and
diameter.  This method allows us to incorporate two unique
characteristics of asteroid families that both result from an origin
in a catastrophic disruption of a single parent: compositional
consistency and minimal orbital velocity differences.

Asteroid families were first identified as groups of objects that
clustered tightly in orbital element-space by \citet{hirayama1918}
nearly a century ago.  Subsequent work has confirmed that families
originate from the catastrophic breakup of a single parent asteroid
after an impact \citep[see][for a recent review of the current state
  of the field]{cellino09}.  This single mineralogical origin causes
families to cluster tightly not only when comparing orbital elements
but also when investigating colors \citep{ivezic02,parker08},
reflectance spectra \citep[e.g.][]{binzel93,cellino01} and albedo
(Mas11).  We note that while the asteroid (4) Vesta shows an albedo
range of $0.10-0.67$ across its surface \citep{reddy12}, and thus any
other differentiated asteroid may have similar large variation, the
Vesta asteroid family has a albedo distribution comparable in width to
other asteroid families (Mas11).

Building further on the origin of families as a result of
collisionally-driven breakups, the size-frequency distribution (SFD)
of asteroid family members acts as a tracer of the physical properties
of the original parent body and can even be used to constrain the
impact velocity and angle \citep{zappalaAIII,durda07}.  However, a
major deficiency in the field to date has been the lack of measured
diameters for the family members, forcing these values to be estimated
based on the apparent visible magnitude of the object.  Furthermore,
the strong selection effects imposed by visible light surveys against
the discovery of low albedo objects (particularly the smallest low
albedo objects) results in a skew in the size distributions of linked
families.  Albedo measurements of the largest bodies in a family are
often available from the Infrared Astronomical Satellite ({\it IRAS})
data set \citep{tedesco02} and can be used to assume an albedo for all
family members, but this can add a significant and systematic error to
the diameters, especially in the cases where it is unclear if the
largest body in a family is indeed associated with the other members
\citep[e.g.][]{cellino01,masiero12bap}.  From the NEOWISE survey we
now have measurements of diameters for over $130,000$ Main Belt
asteroids with relative errors of $\sim10\%$ \citep[see Mas11
  and][]{mainzer11cal}.  Using this new data set along with the
associated proper orbital elements for these objects, we identify the
high-confidence associations of asteroid families detected by NEOWISE.

\section{Data}
\label{sec.data}

Proper orbital elements, which are key to the determination of
asteroid family membership, are the time averaged values of the
semimajor axis, eccentricity, and inclination after removing the
short-period perturbations by Jupiter and Saturn and averaging over
long-period variations \citep{milani98}.  Proper elements are
preferred to osculating elements for family identification as they are
stable over long time periods, and thus asteroid families cluster more
tightly in proper orbital element-space \citep{carpino86}.  We use the
proper orbital elements computed following \citet{milani94,milani98}
that are provided on the AstDys
website\footnote{http://hamilton.dm.unipi.it/astdys/index.php} to
identify asteroid families in this work.

There are two primary methods for determining proper elements:
analytic and synthetic.  Using the analytic method, a Fourier
expansion of the Hamiltonian is solved for directly.  Conversely the
synthetic method integrates the present day osculating orbits using
numerical simulations and determines the proper elements from that
evolution over time.  While the analytic method produces more accurate
results, it has a fundamental limit of $\sim18^\circ$ inclination
above which the solutions degrade due to the truncation of the
Hamiltonian that is typically employed \citep{milani94}.  As NEOWISE
observed the entire sky, it detected and discovered over $15000$
objects at inclinations $>18^\circ$, over ten percent of our total
sample.  In order to consider all asteroids detected by NEOWISE we use
the synthetic proper elements for this work.  This allows us to have
the maximum sample size however synthetic proper elements have their
own inherent limitations: oscillations with periods much longer than
the integration time (typically $\sim10$ million years) will not be
removed properly and the forest of weak secular resonances in the Main
Belt can result in chaotic cases that show up as banding in the
structure of the semimajor axis distribution \citep{knezevic00}.

Following the methods of \citep{knezevic00} we have integrated the
orbits of all Main Belt asteroids that were detected by NEOWISE and have
measured diameters and albedos, but not represented in the AstDys
catalog.  This will include objects with very short arcs, in
particular those discovered by NEOWISE that have had minimal ground-based
followup and thus do not have measured orbits with sufficient quality
to accurately integrate their positions over millions of years.  We
include these for completeness, but with the appropriate caveats
for our results.

We draw our physical properties for Main Belt asteroids primarily from
the measurements made by the WISE spacecraft as part of the NEOWISE
project.  We include in our analysis MBAs that were detected and
discovered throughout the entire mission, both the cryogenic and
post-cryogenic surveys \citep{mainzer11nw,mainzer12pc}.  WISE surveyed
the sky simultaneously in four thermal infrared bands ($3.4~\mu$m,
$4.6~\mu$m, $12~\mu$m, and $22~\mu$m) from a polar low-Earth orbit,
progressing one degree per day.  WISE imaged the entire static sky
over the course of $6$ months starting 14 January 2010, and began a
second pass survey until the exhaustion of the outer cryogen tank on 6
August 2010.  At this point, the longest wavelength channel was lost
and WISE carried out a 3-Band Cryo survey until 29 September 2010 when
the inner cryogen tank was exhausted.  The NEOWISE Post-Cryo Survey to
complete coverage of the largest MBAs and discover new near-Earth
objects began 30 September 2010 and ended on 1 February 2011, using
only the two shortest wavelength bands.  All phases of the mission
employed the WISE Moving Object Processing System (WMOPS) to detect
moving objects in the Level 1 WISE images.  WMOPS required a minimum of
5 detections to link a track (although a typical track had $10-12$
detections), which was then submitted to the Minor Planet Center (MPC)
for further verification.  In total, WMOPS detected $>158,000$ solar
system objects, the majority of which were MBAs.

Diameters and albedos for MBAs seen during the Fully Cryogenic survey
were given in Mas11, while physical properties for objects detected in
the 3-Band Cryo and Post-Cryo Surveys were presented in
\citet{masiero12pc}.  We also include diameters and albedo measured
for $150$ objects by the IRAS mission \citep{tedesco02}.  In most
cases these objects were so large that they saturated in the WISE
images, though a small number were missed due to the exhaustion of
cryogen, observing geometry effects, background contamination, or
other filtering in the WISE or NEOWISE data processing pipelines.
When these data are combined with the synthetic proper orbital
elements our total sample size is $112,286~$MBAs, which we use for the
study presented here.  We note that this is smaller than the total
number of objects detected by NEOWISE due to the fact that
$\sim24,000$ objects do not have stable proper orbital element
solutions.  Nearly three quarters of the unstable objects have
observational arcs shorter than one month implying that their orbits
are not well known and thus their proper elements cannot be reliably
computed.  The remaining $\sim6000$ unstable objects have orbital
elements that are indistinguishable from other MBAs, and orbital arcs
typically of one year or more.  We note that only nine of these
objects have received number designations, and thus the provisional
orbits may still be uncertain, which could result in non-converging
proper elements.  Strangely, we also find that low albedo objects
dominate these long-arc unstable MBAs ($\sim90\%$ have albedos of
$p_V<0.11$), which is much larger than the fraction of low albedo
objects in any of the regions of the Main Belt (Mas11).  Albedo is not
considered when calculating asteroid proper elements and thus should
not affect these results, however low albedo objects are more likely
to be fainter than a high albedo object during a given observation
epoch.  Thus we may be seeing low signal-to-noise uncertainties in a
number of individual observations propagating into the final orbits.
Conversely, these objects may indeed be unstable and thus transitional
in their current orbits, potentially representing low albedo objects
from more distant regions of the Solar system that have been implanted
in the Main Belt.  Distinguishing between these two possibilities,
however, is beyond the scope of the current work and will be the
subject of future investigation.

\begin{figure}[ht]
\begin{center}
\includegraphics[scale=0.7]{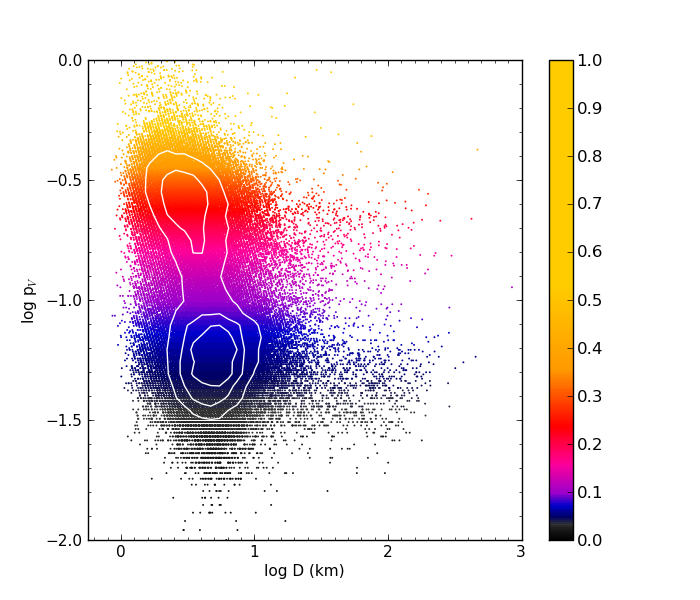}
\caption{Asteroid albedo ($p_V$) vs diameter for the $112,286$ MBAs
  used in this work.  The color of the points also indicates albedo,
  as given in the color bar.  The white contours indicate the density
  of points in the saturated regions.  Two albedo components are
  apparent, with the shift in the low albedo distribution to larger
  diameters a result of the selection bias against small, dark objects
  in optical observations.  The horizontal picket fence effect at low
  albedos is an artifact of the precision of the quoted albedos.  The
  subtle diagonal linear trends at small ($D<2~$km) diameters are an
  artifact of the precision of the literature $H$ absolute magnitudes.
  Albedos larger than $\sim0.5$ are likely artifacts of bad $H$ or $G$
  values \citep[c.f.][]{pravec12}}
\label{fig.albD}
\end{center}
\end{figure}

We show in Figure~\ref{fig.albD} the distribution of albedo as a
function of diameter for all objects used in this work.  Two
populations are apparent, a low-albedo population at $p_V\sim0.06$ and
a high-albedo population at $p_V\sim0.25$ (c.f. Mas11).  The
low-albedo population is shifted to larger diameters compared to the
high-albedo population due to selection biases in optical catalogs.
While the sensitivity of NEOWISE was effectively albedo-independent
\citep{mainzer11neo,grav11troj,grav11hilda}, followup observations of
NEOWISE discoveries by ground-based optical surveys suffer from a
decreased sensitivity to smaller, lower albedo asteroids.  This will
result in the apparent shift in the low albedo population to larger
diameters.  Though we assume for purposes of this analysis that
families have uniform albedos, this bias means that the derived
membership lists for lower albedo families will be missing more small
members than would a high albedo family in the same region of the Main
Belt.

\clearpage

Recently, \citet{pravec12} have presented evidence of possible biases
in the various catalogs of asteroid absolute magnitudes ($H$), and
they highlight the effect that these biases have on the albedo values
derived for a small sample NEOWISE-observed MBAs.  While it is clear
that in many cases the published absolute magnitudes do not reflect
the true values when measured independently, these deviations show a
non-linear relationship, with a maximal deviation at $H\sim14~$mag.
Additionally, the scatter in magnitude difference at $H>6~$mag is
large, and almost always comparable to the mean of the differences.
As such, we chose not to implement a blanket offset correction to the
catalog $H$ values, which would tend to offset the NEOWISE-derived
albedos.  We note that diameters derived from thermal infrared
measurements are largely unaffected by offsets in $H$, and quoted
albedo errors include estimated uncertainties on $H$ and $G$ of $0.3$
and $0.1$ respectively.  Large surveys recently begun or soon to come
online (e.g. Pan-STARRS, LSST) should greatly improve the $H$ catalog
values for most asteroids.  Additionally, implementation of the
$H$-$G_1$-$G_2$ magnitude phase function \citet{muinonen10} may also
improve the determined albedos for cases where there is sufficient
photometric data.  For this work, we use the albedos as published that
are derived from the literature $H$ magnitudes, but with appropriate
error bars.  Future work will revise the catalog of NEOWISE-measured
diameters and albedos using the updated NEOWISE science data
processing system and the most current values of $H$.

\section{Hierarchical Clustering Method}

Asteroid families were originally identified as pairs or groups of
objects with orbital elements that clustered more tightly than would
otherwise be expected from a random distribution of objects.
\citet{zappala90} present a method for association of asteroid
families based on their proper orbital elements called the
Hierarchical Clustering Method (HCM) which uses a distance function
and a velocity cut to link objects together into clusters.  Iterating
over each body in the population, the distance between it and all
other asteroids is calculated by converting the difference in orbital
elements into a pseudo-velocity.  All objects within a given velocity
threshold are added into the family.  Each family member is then
similarly tested, accreting nearby objects into the family, until no
further objects are added.  The selected velocity cutoff will strongly
dictate the size of the family and the reliability of the
associations: a cutoff velocity that is too low will only identify the
core regions of the densest families, while a cutoff velocity that is
too large will include a large number of interlopers in the family
lists or accept spurious groupings of objects that do not have a real
collisional origin.

Following \citet{zappala94}, we apply a distance metric of
\[d=2 \pi F \sqrt{\frac{\frac{5}{4}\left(\frac{a_i-a_c}{a_c}\right)^2 + 2 (ecc_i-ecc_c)^2 + 2 (\sin{inc_i}-\sin{inc_c})^2}{a_c}} \]
where $d$ is the distance in m/s between the objects, $F$ is the
conversion factor changing AU/yr to m/s, $a$, $ecc$, and $inc$ are the
proper semimajor axis, eccentricity, and inclination of the body, and
the subscripts $c$ and $i$ indicate the center body and the body being
tested, respectively.  We perform HCM analysis on every object in our
MBA sample, at velocity limits ranging from $5~$m/s to $200~$m/s in
steps of $5~$m/s.  This allows us to build up a database of all
associations for each object at a range of velocities.  

Following \citet{nesvornyPDS} we divide the Main Belt into three
regions separated by strong Jupiter mean motion resonances: the
inner Main Belt (IMB, $1.8<a<2.5~$AU), the middle Main Belt (MMB,
$2.5<a<2.82~$AU), and the outer Main Belt (OMB, $2.82<a<3.6~$AU).
We set a limit on perihelion distance of $q>1.666~$AU to ensure
Mars-crossing asteroids are not included in our data set.  We use
physical properties as a discriminant in family identification by
dividing each region of the Belt into two groups by albedo.  While the
majority of MBAs show a bimodal albedo distribution
(Figure~\ref{fig.albD}), a small number of asteroid families have a
mean albedo that falls in between these two peaks (Mas11).  In order
to ensure that these families can be properly identified and that the
wings of the albedo distribution of each family are not truncated, we
have separated the regions by albedo allowing for an overlap region in
between.  As such, objects with moderate albedo will appear in both
lists.  Our high albedo group for each Main Belt region includes all
objects with $p_V>0.065$ while the low albedo group includes objects
with $p_V<0.155$, which represents a buffer of $\pm40\%$ relative to
the central minimum reflectance of the MBA albedo distribution at
$p_V=0.11$ (Mas11).  The total number of objects in each group
searched as well as the bounding semimajor axes and albedo are shown
in Table~\ref{tab.regions}.

\begin{table}[ht]
\begin{center}
\caption{Main Belt regions used for HCM analysis}
\vspace{1ex}
{\scriptsize
\noindent
\begin{tabular}{ccccc}
\tableline
Region & Semimajor Axis Range (AU) & Albedo Range & Number$^\dagger$ & QRL$^\ddagger$ (m/s)\\
\tableline
IMB$_{high}$ & $1.8<a<2.5$  & $p_V>0.065$ & 21013 & $110 \pm 6$\\
IMB$_{low}$  & $1.8<a<2.5$  & $p_V<0.155$ & 10622 & $132 \pm 7$\\
MMB$_{high}$ & $2.5<a<2.82$ & $p_V>0.065$ & 26214 & $101 \pm 8$\\
MMB$_{low}$  & $2.5<a<2.82$ & $p_V<0.155$ & 22958 & $102 \pm 7$\\
OMB$_{high}$ & $2.82<a<3.6$ & $p_V>0.065$ & 24204 & $102 \pm 7$\\
OMB$_{low}$  & $2.82<a<3.6$ & $p_V<0.155$ & 38691 & $107 \pm 8$\\
\hline
\end{tabular}
\\
{\scriptsize $^\dagger$Summed number of objects in all regions is greater than total population due to overlap between albedos; see text for details\\$^\ddagger$Quasi-Random Level used for determining significant family linkages}
}
\label{tab.regions}
\end{center}
\end{table}

To enable rapid searching of each region investigated we employ a
k-dimensional (KD)-tree query to perform an initial reduction in the possible
associations for each object.  KD-trees are computational methods of
dividing up multidimensional data to increase the efficiency of
searches for specific data points within that space.  All objects
identified by the KD-tree test are then compared to the velocity limit
of that run to test for family membership.  This process provides a
dramatic reduction in run time of the search procedure.

After identifying the family associations at each velocity cut for
each region of the Belt, we need to determine which velocity cut
represents the optimal blend between completeness and accuracy.  We
follow \citet{zappala94} and use a Quasi-Random Level (QRL) test to
determine at which point background objects begin to become a
significant contributor to family lists.  The methodology behind a QRL
test is to construct a synthetic population based on the orbital
elements of the real MBAs and determine at which velocity limit the
synthetic objects begin to be linked by HCM.  This velocity can then
be considered the level at which a quasi-random population of objects
begin to contribute to family lists; any linkages below this limit are
unlikely to be random associations.  We note that as the overall
population of known MBAs grows, the average distance between any two
asteroids will decrease and thus the QRL will get smaller over time
(although as more family members are observed the velocity cut at
which they link will also shrink).  We use a lower limit of ten
objects as the minimum size group we consider both for family
membership and QRL determination.

In order to construct a representative synthetic background
population, one must know which objects are in the background as
opposed to belonging to a family and thus {\it a priori} have a list
of known family members.  This becomes especially critical when a large,
dense family dominates a small area of the phase space.  If these
families are not removed, the QRL will be made artificially smaller as
real family members begin to link together in the quasi-random
population.  

Removing a too many objects will decrease the number density and thus
increase the calculated QRL, while removing too few will have the
opposite effect.  Similarly, replacing removed objects with randomly
generated ones will artificially increase the number of objects not
associated with families when compared to the real population, causing
them to link at a lower level than the observed background.  The two
parameters determining the number of objects removed are the linking
velocity level chosen and the size of the smallest family removed.  By
varying these two parameters, we tested their effect on the final QRL
determination for all regions of the Main Belt.  We find that the size
of the smallest family removed (from $50-200$ members) has only a
minimal effect on the determined QRL, as it is the few largest
families that dominate the artificially reduced QRL levels.  On the
other hand, the velocity level chosen (from $75-150~$m/s) does have a
strong effect on the final determined QRL, which track each other
closely.  We use the lists of members of the largest families
identified by \citet{nesvornyPDS} to find a velocity cut in our
determination that most closely reproduces those families, and use
that velocity level as the limit for large family removal.

In this implementation we use limits of 100 members and a velocity cut
of $125~$m/s to remove the largest families.  Any asteroid in one of
these families was removed from the sample used to generate the
quasi-random population.  We then divided each region into subregions
by semimajor axis so that each eccentricity-inclination slice
contained $10\%$ of the total number of objects in the entire region.
The quasi-random population was built by randomly drawing $a$, $ecc$,
and $inc$ values from the members of the slice.  This ensured that the
distribution of these three orbital parameters remained identical for
the original and quasi-random populations, minus the removed large
families.  The slices were then reassembled and HCM was run on this
population for all velocity cuts from $5-200~$m/s is steps of $5~$m/s.
This process was performed ten times for each region of the Main Belt,
and the velocity levels of the first five pseudo-families of each
trial were averaged to determine the QRL for the region.  This
provides both a mean QRL level for each region as well as an estimate
of the uncertainty on that value, both of which are shown in
Table~\ref{tab.regions}.

\begin{sidewaysfigure}[ht]
\begin{center}
\includegraphics[scale=0.35]{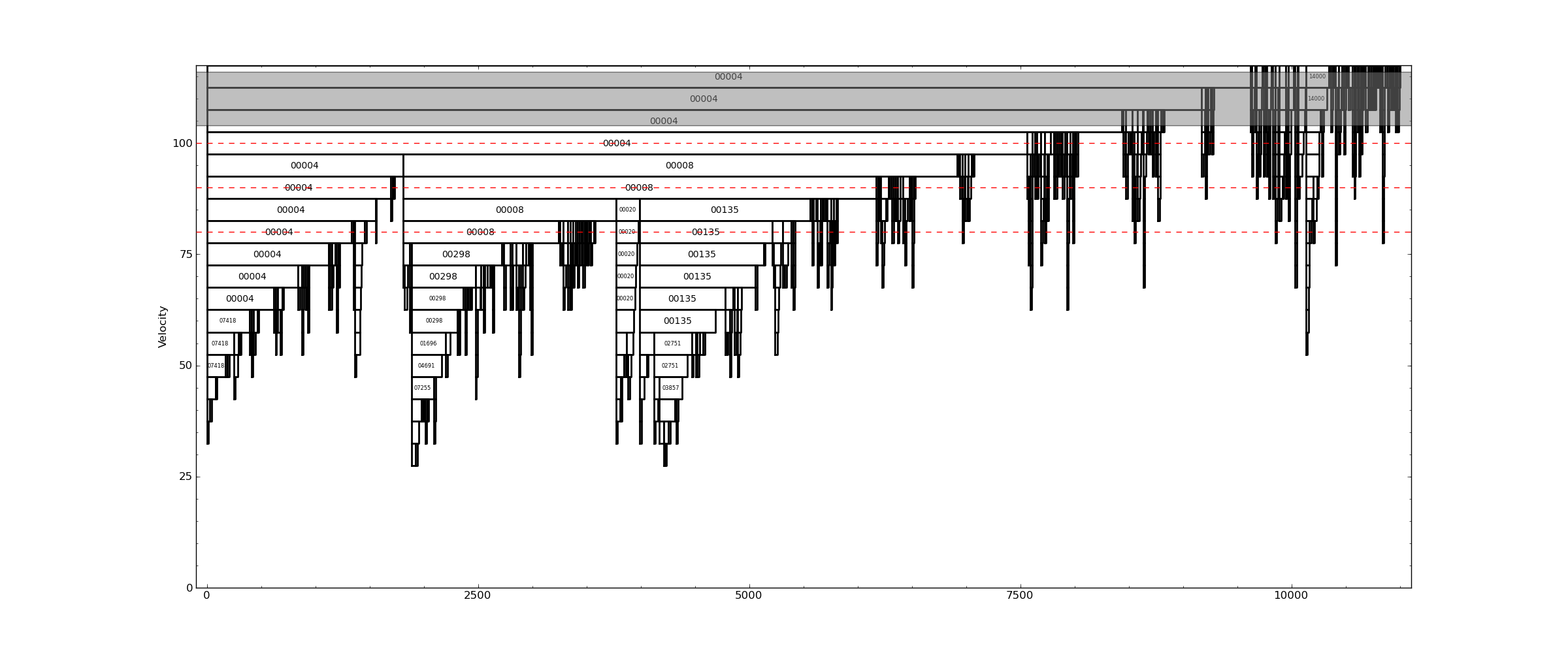}
\protect\caption{Stalactite diagram showing family associations for
  the high albedo objects in the inner Main Belt.  Families are
  identified by the designation number of the largest asteroid
  contained in the family list and length of each bar indicates the
  number of objects contained in that family at that velocity cut.
  The count at the bottom of the plot is an arbitrary sum of the bars
  at the topmost level.  Small families are left unlabeled for clarity
  but are discussed in the text.  The grey box shows the $1\sigma$
  range around the QRL level: family associations within this range
  are likely contaminated by background objects.  The three dashed red
  lines indicate the velocities used to extract family membership
  lists.}
\label{fig.imbh_stal}
\end{center}
\end{sidewaysfigure}

\begin{sidewaysfigure}[ht]
\begin{center}
\includegraphics[scale=0.35]{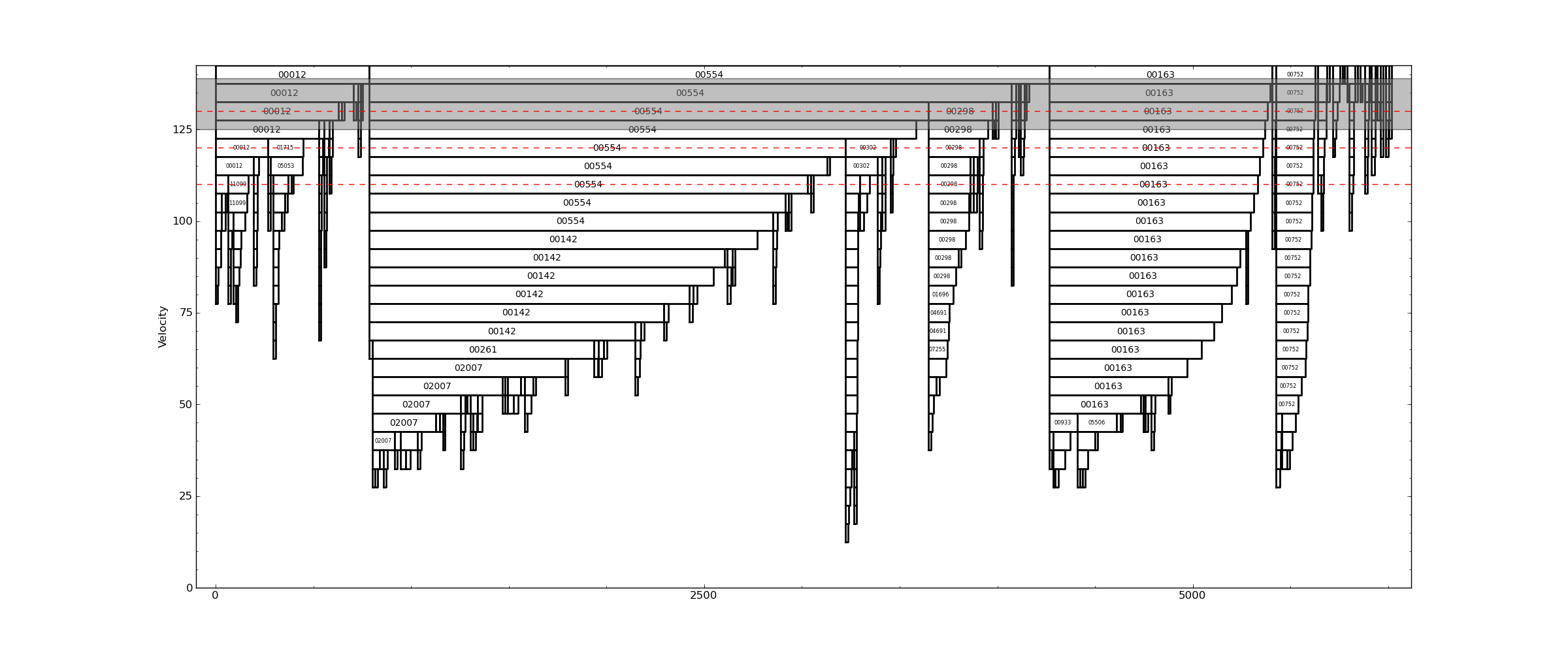}
\protect\caption{The same as Figure~\ref{fig.imbh_stal} but for the
  low albedo inner Main Belt asteroids.}
\label{fig.imbl_stal}
\end{center}
\end{sidewaysfigure}

\begin{sidewaysfigure}[ht]
\begin{center}
\includegraphics[scale=0.35]{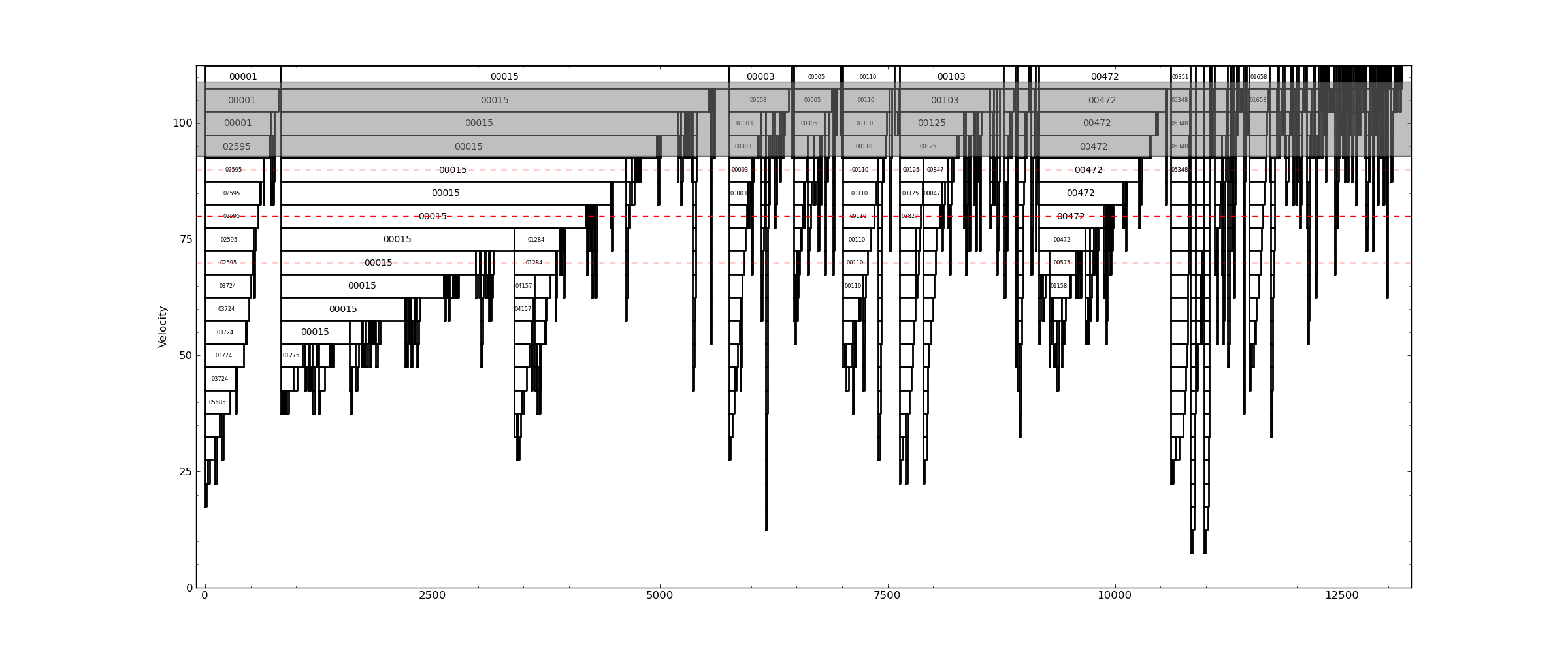}
\protect\caption{The same as Figure~\ref{fig.imbh_stal} but for the
  high albedo middle Main Belt asteroids.}
\label{fig.mmbh_stal}
\end{center}
\end{sidewaysfigure}

\begin{sidewaysfigure}[ht]
\begin{center}
\includegraphics[scale=0.35]{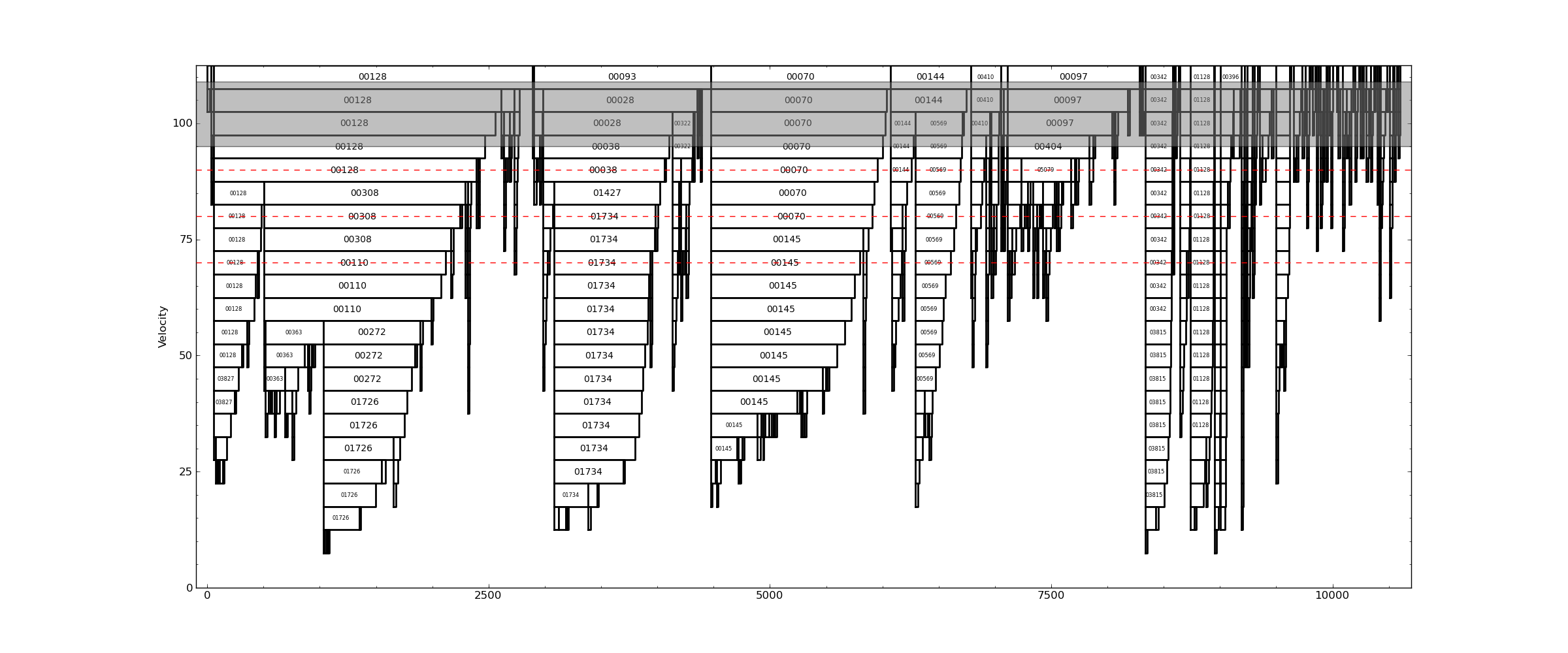}
\protect\caption{The same as Figure~\ref{fig.imbh_stal} but for the
  low albedo middle Main Belt asteroids.}
\label{fig.mmbl_stal}
\end{center}
\end{sidewaysfigure}

\begin{sidewaysfigure}[ht]
\begin{center}
\includegraphics[scale=0.35]{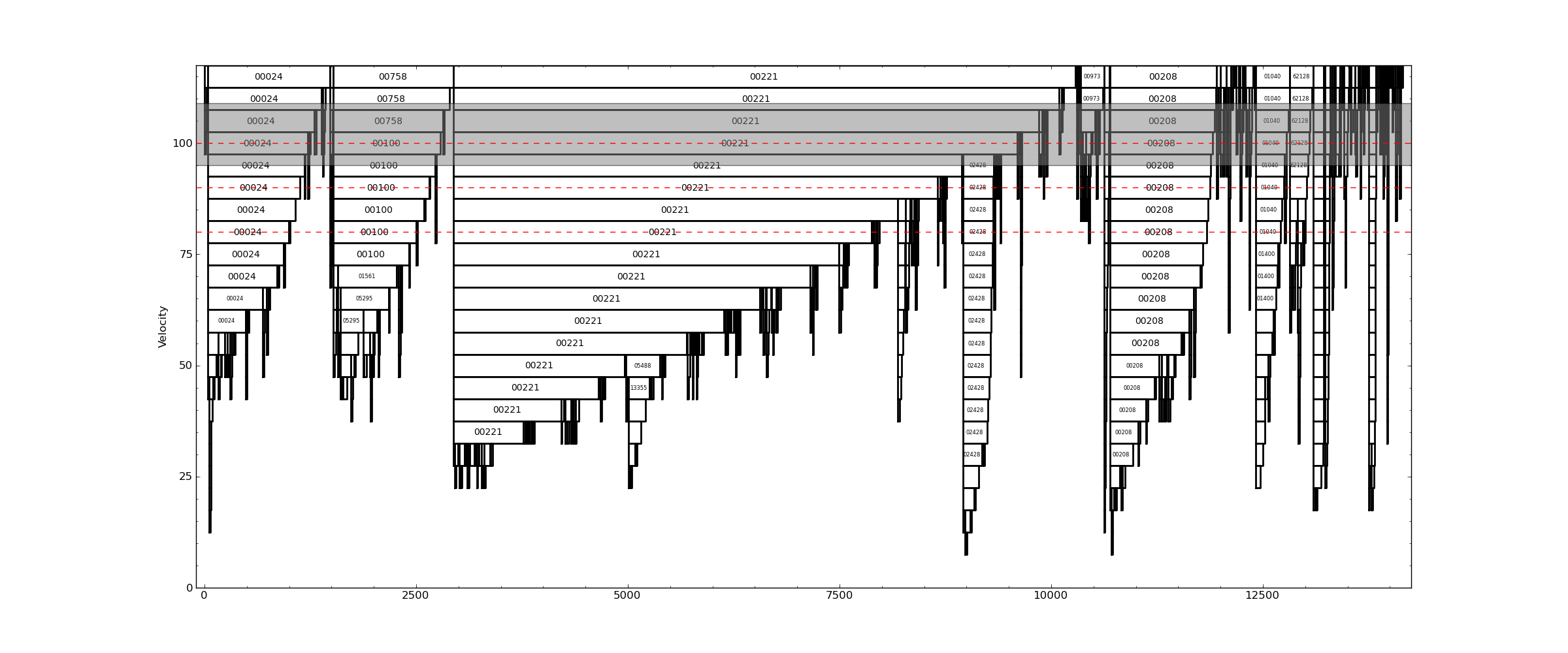}
\protect\caption{The same as Figure~\ref{fig.imbh_stal} but for the
  high albedo outer Main Belt asteroids.}
\label{fig.ombh_stal}
\end{center}
\end{sidewaysfigure}

\begin{sidewaysfigure}[ht]
\begin{center}
\includegraphics[scale=0.35]{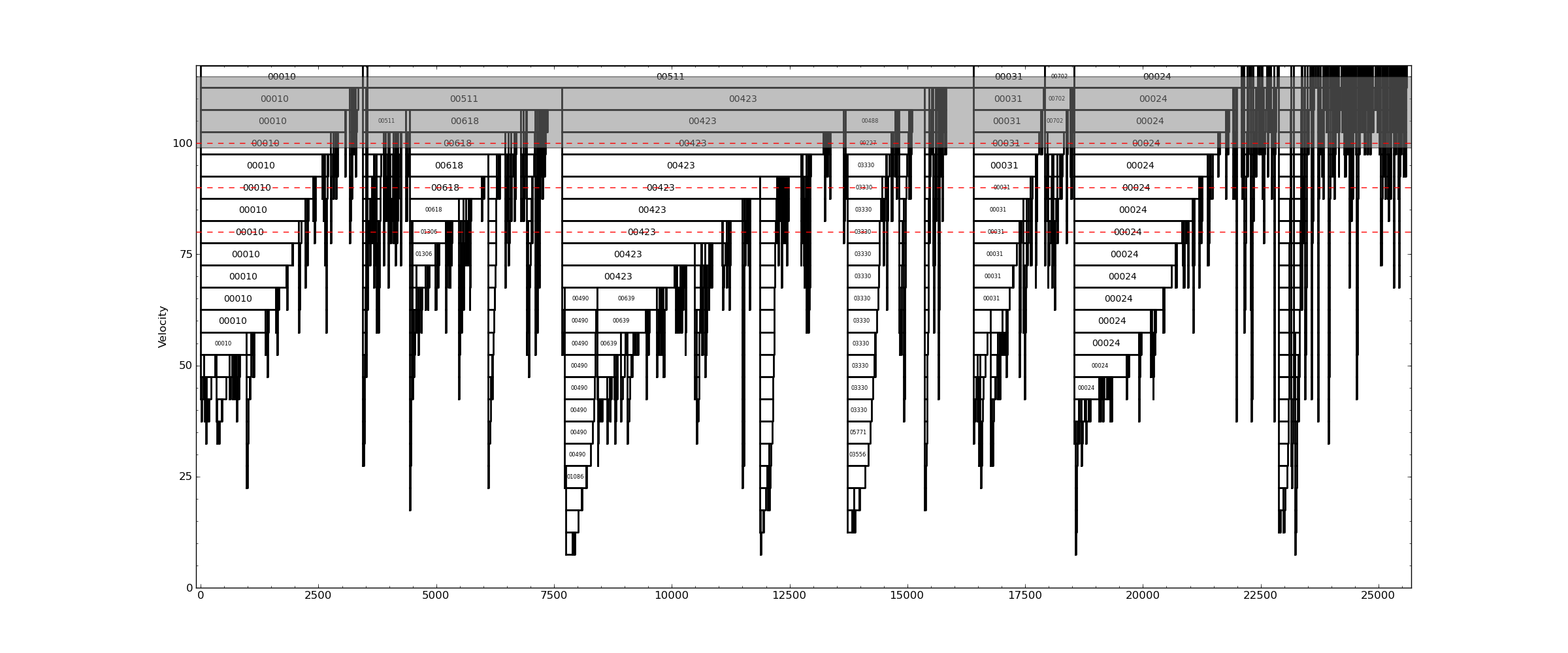}
\protect\caption{The same as Figure~\ref{fig.imbh_stal} but for the
  low albedo outer Main Belt asteroids.}
\label{fig.ombl_stal}
\end{center}
\end{sidewaysfigure}

A common way of representing asteroid families is a `stalactite' plot,
which shows the membership of each family as a function of cutoff
velocity as well as which families merge or fragment at various cutoff
levels.  We show stalactite plots for each of the six regions
considered here in Figures~\ref{fig.imbh_stal}-\ref{fig.ombl_stal}.
We also show the $1\sigma$ range of the QRL as the grey box overlaid
on the stalactites.  At each velocity step, the families are labeled
with the numerical designation of the member with the largest measured
diameter.  Designations will change with decreasing velocity as
objects connected more loosely fall out of the family list.  We have
left the smallest families on each plot unlabeled to preserve the
clarity of the figures.

\clearpage

Stalactite plots are read from top to bottom, showing the least- to
most-significant clusters of objects.  For example, in
Figure~\ref{fig.imbh_stal} we can follow the evolution of the objects
linked to (8) Flora.  At large velocities ($v_{link}>100~$m/s) a large
fraction of the asteroids in the IMB$_{high}$ region link to (4)
Vesta.  Even below the QRL level at $v_{link}=100~$m/s all of the
large families remain grouped together.  At $v_{link}=95~$m/s the
Vesta and Flora clumps separate, and at $v_{link}=85~$m/s the Flora
clump breaks up into the Flora, (20) Massalia, and (135) Hertha
families.  At velocities below $v_{link}=80~$m/s Flora no longer links
with the family, and nearly half of the other family members are
rejected at this cutoff as well.  In contrast, Vesta remains linked to
its family down to $v_{link}=65~$m/s, though this family also loses a
substantial fraction of linked members as the cutoff velocity
decreases.

\section{Results and Discussion}

\subsection{Identified Families}
\label{sec.families}

We use our stalactite plots as a first cut to guide our selection of
the optimal cutoff velocity at which to extract each family list.  For
each region we have extracted families at three different velocity
cuts: one at or near the lower edge of the identified QRL zone and two
others at $10~$m/s and $20~$m/s below that level to separate
overlapping families (e.g. Vesta and Flora) and winnow out background
objects that may be connecting below the QRL level.  This is similar
to the method employed by \citet{milani10} to identify the Hungaria
family.  The extraction levels for each region are indicated in
Figures~\ref{fig.imbh_stal}-\ref{fig.ombl_stal} by the dashed red
lines.  We note that due to the assumptions inherent in our QRL
determination, some of the associations presented here may be
incomplete or even spurious.  A more detailed analysis of each family
individually can refine these associations, and will be the subject of
future work.

\begin{figure}[ht]
\begin{center}
\includegraphics[scale=0.6]{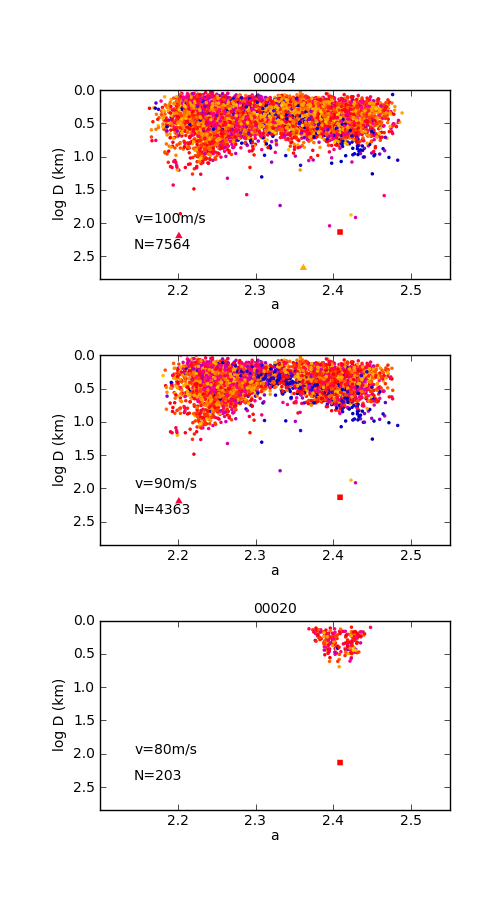}
\caption{Diameter vs proper semimajor axis for the Massalia family at
  three different cutoff velocities (from top to bottom: $100~$m/s,
  $90~$m/s, $80~$m/s).  The color of each point indicates its albedo
  as in Figure~\ref{fig.albD}, while $N$ lists the number of objects
  linked at that cutoff.  Each subplot is titled with the largest
  object included in the family, the presumed parent.  Here, the
  family of (20) Massalia (marked as a square) blends with the
  families of (8) Flora and (4) Vesta (both marked as triangles) at
  velocity cutoffs larger than $80~$m/s.}
\label{fig.massalia}
\end{center}
\end{figure}

\begin{figure}[ht]
\begin{center}
\includegraphics[scale=0.6]{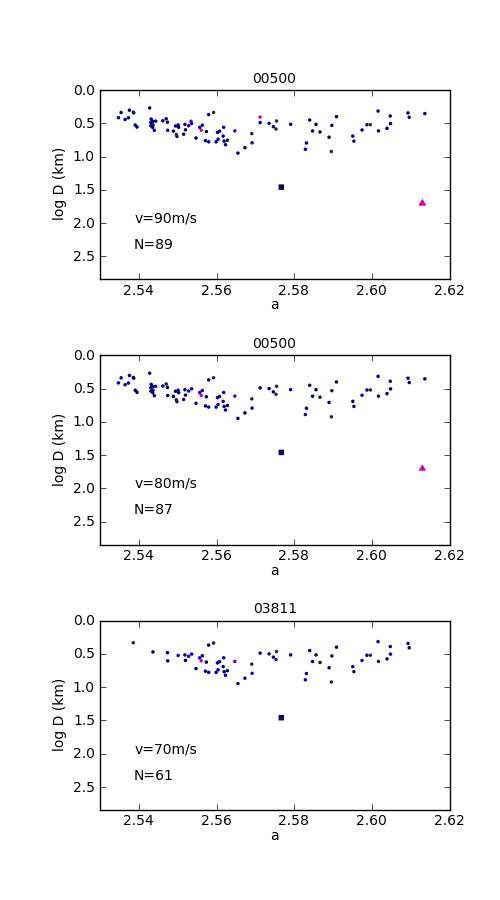}
\caption{Diameter vs proper semimajor axis for the Karma family at
  three different cutoff velocities (from top to bottom: $90~$m/s,
  $80~$m/s, $70~$m/s).  The color of each point indicates its albedo
  as in Figure~\ref{fig.albD}, while $N$ lists the number of objects
  linked at that cutoff.  Each subplot is titled with the largest
  object included in the family, the presumed parent.  Here, (500)
  Selinur (marked as a triangle) links with the family at larger
  velocities, but is rejected at $70~$m/s, making (3811) Karma (marked
  as a square) the most likely parent. }
\label{fig.karma}
\end{center}
\end{figure}

By plotting the diameter of each family member against their proper
semimajor axis, and color-coding the points to represent albedo, we
construct ``petal'' plots which can be used to diagnose the
reliability of the family association.  An ideal family will have a
large parent body at the bottom-center of the plot with the family
spreading to smaller diameters and larger distance in semimajor axis
from the parent, as would be expected from a family evolved by
Yarkovsky drift \citep{bottke06}.  For families in dense regions of
the Main Belt, large families will link together even at levels below
the QRL.  By choosing a lower velocity cutoff we can disentangle these
overlapping families.  Figure~\ref{fig.massalia} shows an example of
this, where the Massalia family blends with the Flora and Vesta family
as the velocity cutoff is increased.  We can also use this technique
to filter out large background objects that may be linked to a family
and thus mis-identified as the parent body.  Figure~\ref{fig.karma}
shows an example of this scenario, where (500) Selinur is linked to
the family at larger velocities, but at the lowest cutoff is rejected
from the family.  The albedo of (500) Selinur and its placement on the
plots indicate it is unlikely to be the parent of the family, and that
(3811) Karma is a much more likely parent body.  This results in a
family with a very classic petal shape.

\clearpage

Using the physical and orbital parameters of the objects that were
linked together at each velocity cutoff, we identify the highest
reliability families in the Main Belt that can be detected with this
technique.  To ensure a high level of reliability and measure
statistically significant family properties we have only included
families with $>40$ members.  We find $76$ families throughout the
Main Belt that pass these cutoffs representing $38,298$ MBAs
(approximately $35\%$ of all objects considered), and another $\sim60$
candidate groupings below of $40$ member limit that are too small to
be definite detections.  We give a list of the identified families,
with the number (in MPC-packed format) and name of the largest member,
the HCM velocity the family was extracted at ($v_{link}$), the median
proper semimajor axis, eccentricity, and inclination for the family
($a_{med}$, $ecc_{med}$, $inc_{med}$, respectively), and the number of
linked family members (N) in Table~\ref{tab.stats}.  We also include
average physical properties for each family in Table~\ref{tab.stats} (see
Section~\ref{sec.phys} for description of these parameters).  Names
with a `*' suffix indicate cases of ambiguous parent bodies (see
below).  In Table~\ref{tab.list} we present the full list of asteroids
associated with families, their orbital and physical parameters, and
identify the family to which they have been linked.

\begin{center}
{\scriptsize
\begin{longtable}{rlccccccccccr}
\caption{Orbital elements, median and maximum diameters, average albedos, and raw SFD slopes ($\alpha$) for observed asteroid families}\label{tab.stats}\\
\hline\hline
Number &  Name  &  $v_{link}$ & $a_{med}$  &  $ecc_{med}$ & $inc_{med}$ &  $D_{max}$ &  $D_{med}$ & $\overline{p_V}$ & $\sigma_{p_V}$ &  $\alpha_{SFD}$ & $\sigma_{\alpha}$&   N\\
 &   &  (m/s) & (AU)  &   & (deg) & (km) & (km) &  &  &   &  &   \\
\hline\hline
\endfirsthead
\caption[]{(continued)}\\
\hline\hline
Number &  Name  &  $v_{link}$ & $a_{med}$  &  $ecc_{med}$ & $inc_{med}$ &  $D_{max}$ &  $D_{med}$ & $\overline{p_V}$ & $\sigma_{p_V}$ &  $\alpha_{SFD}$ & $\sigma_{\alpha}$&   N\\
 &   &  (m/s) & (AU)  &   & (deg) & (km) & (km) &  &  &   &  &   \\
\hline\hline
\endhead
\hline
\endfoot
00004 &           Vesta &  80 & 2.3469 & 0.0972 &  6.6786 & 468.30 &   2.50 &  0.361 &   0.111 &   -3.448 &   0.038 &   1331\\
00008 &           Flora &  80 & 2.2543 & 0.1409 &  5.4513 & 155.74 &   2.70 &  0.288 &   0.088 &   -2.589 &   0.032 &    929\\
00020 &        Massalia &  80 & 2.4035 & 0.1636 &  1.4211 & 135.68 &   1.91 &  0.243 &   0.066 &   -3.929 &   0.164 &    203\\
00135 &          Hertha &  80 & 2.3990 & 0.1796 &  2.4243 &  82.15 &   2.42 &  0.284 &   0.091 &   -3.375 &   0.050 &   1113\\
00254 &        Augusta* &  90 & 2.1986 & 0.1213 &  4.1490 &  11.85 &   2.45 &  0.305 &   0.094 &   -1.727 &   0.111 &     72\\
00434 &        Hungaria & 100 & 1.9466 & 0.0771 & 21.0020 &   8.93 &   1.66 &  0.722 &   0.156 &   -2.102 &   0.397 &     48\\
00587 &       Hypsipyle & 100 & 2.3354 & 0.2203 & 24.0667 &  12.23 &   3.01 &  0.318 &   0.093 &   -2.285 &   0.389 &     43\\
01646 &      Rosseland* & 100 & 2.3501 & 0.0983 &  8.0594 &  12.47 &   3.23 &  0.194 &   0.047 &   -2.393 &   0.293 &     46\\
02409 &         Chapman &  80 & 2.2746 & 0.1342 &  3.2275 &   8.70 &   2.31 &  0.288 &   0.076 &   -2.391 &   0.205 &     78\\
04689 &            Donn &  90 & 2.2763 & 0.1166 &  4.7007 &   6.14 &   2.72 &  0.278 &   0.074 &   -3.326 &   0.351 &     60\\
13698 &          13698* &  80 & 2.4368 & 0.1129 &  6.3787 &   5.97 &   2.75 &  0.359 &   0.107 &   -4.185 &   0.309 &     87\\
00012 &           Klio* & 120 & 2.3859 & 0.1922 &  9.3462 & 126.64 &   3.57 &  0.062 &   0.019 &   -2.750 &   0.082 &    269\\
01715 &           Salli & 120 & 2.4114 & 0.2289 & 10.9966 &  24.16 &   3.49 &  0.062 &   0.021 &   -2.907 &   0.103 &    178\\
00163 &         Erigone & 120 & 2.3718 & 0.2088 &  5.0486 &  81.58 &   2.98 &  0.051 &   0.012 &   -3.229 &   0.040 &   1093\\
00298 &     Baptistina* & 100 & 2.2737 & 0.1451 &  5.6067 &  21.14 &   2.34 &  0.158 &   0.029 &   -2.692 &   0.048 &    549\\
00302 &        Clarissa & 120 & 2.3967 & 0.1082 &  2.6910 &  38.53 &   3.02 &  0.056 &   0.017 &   -3.242 &   0.110 &    228\\
00554 &         Polana* & 120 & 2.3566 & 0.1485 &  2.8258 & 102.78 &   3.11 &  0.057 &   0.015 &   -2.376 &   0.016 &   2438\\
00623 &        Chimaera & 130 & 2.4456 & 0.1500 & 14.8210 &  44.09 &   4.16 &  0.059 &   0.011 &   -2.507 &   0.357 &     46\\
00752 &       Sulamitis & 120 & 2.4407 & 0.0894 &  5.0428 &  60.85 &   3.38 &  0.052 &   0.013 &   -2.408 &   0.090 &    191\\
00003 &            Juno &  80 & 2.6654 & 0.2351 & 13.3530 & 246.60 &   2.15 &  0.252 &   0.062 &   -3.318 &   0.121 &    196\\
00005 &         Astraea &  80 & 2.5823 & 0.1986 &  4.4909 & 113.00 &   2.50 &  0.279 &   0.072 &   -2.962 &   0.178 &     94\\
00015 &         Eunomia &  70 & 2.6214 & 0.1497 & 13.1676 & 299.21 &   4.07 &  0.268 &   0.073 &   -2.958 &   0.025 &   2140\\
00472 &           Roma* &  80 & 2.6022 & 0.0915 & 14.8032 &  47.04 &   3.63 &  0.257 &   0.078 &   -2.147 &   0.024 &    712\\
00480 &           Hansa &  80 & 2.6557 & 0.0111 & 21.9749 &  65.67 &   3.12 &  0.249 &   0.091 &   -2.358 &   0.231 &     65\\
00606 &        Brangane &  70 & 2.5817 & 0.1803 &  9.6134 &  39.53 &   2.87 &  0.112 &   0.034 &   -3.322 &   0.354 &     57\\
00808 &          Merxia &  80 & 2.7431 & 0.1341 &  5.0083 &  37.68 &   3.16 &  0.229 &   0.062 &   -2.624 &   0.180 &     90\\
00847 &           Agnia &  80 & 2.7902 & 0.0723 &  3.8130 &  30.08 &   3.84 &  0.227 &   0.070 &   -2.850 &   0.107 &    180\\
01658 &           Innes &  80 & 2.5799 & 0.1727 &  7.5910 &  13.81 &   3.05 &  0.256 &   0.071 &   -3.284 &   0.159 &    155\\
02595 &    Gudiachvili* &  80 & 2.7718 & 0.1311 &  9.1140 &  14.62 &   3.97 &  0.265 &   0.069 &   -3.087 &   0.055 &    584\\
00539 &          Pamina &  80 & 2.7428 & 0.1628 &  8.2446 &  56.04 &   4.57 &  0.057 &   0.019 &   -2.588 &   0.177 &     88\\
01734 &   Zhongolovich* &  80 & 2.7861 & 0.1958 &  7.8512 &  26.70 &   4.86 &  0.054 &   0.014 &   -2.557 &   0.034 &    903\\
00145 &          Adeona &  70 & 2.6513 & 0.1663 & 11.6481 & 132.59 &   4.46 &  0.059 &   0.014 &   -2.704 &   0.027 &   1321\\
00128 &         Nemesis &  70 & 2.7343 & 0.0896 &  4.8703 & 193.08 &   3.64 &  0.071 &   0.023 &   -3.718 &   0.094 &    390\\
00363 &           Padua &  55 & 2.7344 & 0.0411 &  5.3190 &  86.04 &   4.15 &  0.067 &   0.018 &   -2.655 &   0.048 &    512\\
00272 &        Antonia* &  55 & 2.7842 & 0.0476 &  4.3932 &  25.67 &   4.01 &  0.046 &   0.011 &   -2.966 &   0.040 &    861\\
00144 &         Vibilia &  90 & 2.6715 & 0.1888 &  3.8417 & 142.38 &   4.33 &  0.064 &   0.013 &   -3.095 &   0.123 &    184\\
00322 &           Phaeo &  90 & 2.7875 & 0.1935 &  9.4217 &  73.15 &   3.37 &  0.068 &   0.017 &   -2.956 &   0.282 &     72\\
00342 &       Endymion* &  70 & 2.5719 & 0.1401 &  8.7950 &  64.27 &   3.39 &  0.043 &   0.013 &   -2.812 &   0.086 &    230\\
00396 &          Aeolia &  80 & 2.7399 & 0.1678 &  3.4398 &  37.29 &   2.85 &  0.094 &   0.027 &   -2.823 &   0.289 &     62\\
00404 &        Arsinoe* &  80 & 2.6257 & 0.2304 & 13.1558 & 105.41 &   4.17 &  0.051 &   0.015 &   -2.391 &   0.137 &    113\\
00410 &         Chloris &  90 & 2.7458 & 0.2522 &  8.7834 & 118.93 &   5.33 &  0.084 &   0.029 &   -2.571 &   0.144 &    116\\
03811 &           Karma &  70 & 2.5690 & 0.1066 & 10.7836 &  28.75 &   3.96 &  0.054 &   0.010 &   -2.603 &   0.329 &     61\\
00569 &            Misa &  80 & 2.6479 & 0.1778 &  2.2867 &  78.93 &   3.46 &  0.052 &   0.015 &   -2.447 &   0.054 &    357\\
01128 &          Astrid &  80 & 2.7773 & 0.0484 &  0.6818 &  48.63 &   3.50 &  0.046 &   0.011 &   -2.585 &   0.083 &    201\\
01668 &           Hanna &  80 & 2.7932 & 0.1766 &  4.2265 &  25.83 &   3.86 &  0.051 &   0.013 &   -3.608 &   0.185 &    122\\
02669 &   Shostakovich* &  90 & 2.7687 & 0.1739 &  9.2127 &  16.47 &   5.14 &  0.051 &   0.016 &   -3.001 &   0.203 &     98\\
03567 &         Alvema* &  90 & 2.7702 & 0.2806 &  8.3054 &  14.53 &   4.24 &  0.056 &   0.016 &   -3.302 &   0.296 &     62\\
05079 &        Brubeck* &  90 & 2.5722 & 0.2493 & 12.4977 &  16.95 &   4.00 &  0.066 &   0.018 &   -3.011 &   0.071 &    441\\
00208 &      Lacrimosa* &  90 & 2.8913 & 0.0489 &  2.1090 &  49.99 &   4.85 &  0.238 &   0.062 &   -2.392 &   0.024 &   1175\\
00179 &   Klytaemnestra &  80 & 2.9852 & 0.0674 &  8.7805 &  74.59 &   4.17 &  0.217 &   0.075 &   -5.168 &   0.606 &     90\\
00221 &            Eos* &  90 & 3.0248 & 0.0744 & 10.1660 &  95.63 &   4.98 &  0.157 &   0.045 &   -2.320 &   0.010 &   5718\\
01040 &        Klumpkea &  90 & 3.1230 & 0.1976 & 16.8220 &  22.67 &   4.08 &  0.235 &   0.066 &   -2.934 &   0.091 &    333\\
03985 &      Raybatson* &  90 & 2.8516 & 0.1221 & 15.0523 &  22.11 &   3.54 &  0.167 &   0.056 &   -3.572 &   0.234 &    126\\
00010 &          Hygiea & 100 & 3.1573 & 0.1308 &  5.2384 & 453.24 &   5.75 &  0.068 &   0.022 &   -2.484 &   0.014 &   2757\\
00024 &          Themis & 100 & 3.1371 & 0.1505 &  1.3982 & 193.54 &   6.88 &  0.066 &   0.021 &   -2.177 &   0.012 &   3052\\
00031 &      Euphrosyne & 100 & 3.1679 & 0.1940 & 26.5578 & 281.98 &   6.11 &  0.056 &   0.016 &   -4.404 &   0.053 &   1392\\
00081 &     Terpsichore &  90 & 2.8866 & 0.1848 &  8.1925 & 123.96 &   4.68 &  0.051 &   0.014 &   -3.842 &   0.792 &     49\\
00087 &          Sylvia &  90 & 3.4981 & 0.0558 &  9.8460 & 288.38 &   7.66 &  0.056 &   0.016 &   -2.624 &   0.365 &     60\\
00096 &           Aegle &  90 & 3.0570 & 0.1844 & 16.5379 & 177.77 &   6.63 &  0.071 &   0.016 &   -3.418 &   0.296 &     86\\
03330 &       Gantrisch &  90 & 3.1466 & 0.1974 & 10.1892 &  37.64 &   5.51 &  0.043 &   0.011 &   -3.171 &   0.054 &    734\\
00276 &       Adelheid* & 100 & 3.1953 & 0.0665 & 21.8977 & 100.35 &   8.76 &  0.065 &   0.019 &   -2.244 &   0.050 &    358\\
00283 &            Emma &  80 & 3.0508 & 0.1142 &  9.0734 & 145.55 &   5.74 &  0.044 &   0.013 &   -3.160 &   0.081 &    340\\
00490 &         Veritas &  65 & 3.1700 & 0.0615 &  9.2533 & 118.80 &   5.79 &  0.066 &   0.020 &   -2.767 &   0.044 &    686\\
24649 &      Balaklava* &  90 & 3.1880 & 0.2108 & 14.1109 &  16.79 &   4.23 &  0.056 &   0.011 &   -2.244 &   0.148 &     88\\
00511 &          Davida &  90 & 3.1476 & 0.1903 & 14.4627 & 285.84 &   7.66 &  0.058 &   0.018 &   -1.455 &   0.068 &    104\\
00618 &        Elfriede &  80 & 3.1886 & 0.0582 & 15.8548 & 131.23 &   5.16 &  0.054 &   0.018 &   -3.621 &   0.799 &     45\\
01306 &        Scythia* &  80 & 3.1408 & 0.0896 & 16.4274 &  72.24 &   6.59 &  0.057 &   0.017 &   -2.482 &   0.035 &    705\\
01303 &         Luthera &  90 & 3.2169 & 0.1210 & 18.7869 & 102.43 &   6.81 &  0.047 &   0.013 &   -3.794 &   0.176 &    176\\
00702 &          Alauda & 100 & 3.2184 & 0.0183 & 21.5554 & 196.47 &  10.22 &  0.065 &   0.015 &   -1.747 &   0.197 &     49\\
00778 &       Theobalda &  90 & 3.1745 & 0.2541 & 14.3001 &  55.32 &   6.04 &  0.061 &   0.020 &   -3.199 &   0.156 &    144\\
00780 &         Armenia &  90 & 3.1057 & 0.0690 & 18.1677 & 114.26 &   5.46 &  0.053 &   0.014 &   -4.563 &   1.482 &     40\\
00816 &         Juliana &  90 & 2.9887 & 0.1460 & 13.2853 &  50.08 &   5.62 &  0.042 &   0.014 &   -3.260 &   0.640 &     42\\
00845 &           Naema &  90 & 2.9333 & 0.0355 & 11.9642 &  58.53 &   5.25 &  0.059 &   0.016 &   -3.955 &   0.141 &    246\\
00928 &         Hildrun &  90 & 3.1459 & 0.1940 & 16.4752 &  62.54 &   5.55 &  0.052 &   0.013 &   -3.096 &   0.159 &    111\\
02621 &           Goto* &  80 & 3.0922 & 0.1207 & 12.2630 &  47.92 &   5.64 &  0.080 &   0.034 &   -3.708 &   0.384 &     71\\
01113 &           Katja &  90 & 3.1136 & 0.1324 & 13.7567 &  48.37 &   7.19 &  0.067 &   0.030 &   -2.027 &   0.229 &     51\\
\hline
\end{longtable}
}
\end{center}

\begin{table}[ht]
\begin{center}
\caption{Orbital and physical parameters for Main Belt asteroids
  associated with dynamical families.  Table 3 is published in its
  entirety in the electronic edition of ApJ; a portion is shown here
  for guidance regarding its form and content.}
\vspace{1ex}
\noindent
\begin{tabular}{ccccccccc}
\tableline
 Name  &   $a$ (AU)  &  $ecc$ & $inc$ (deg) & $D$ (km) & $\sigma_D$ & $p_V$ & $\sigma_{p_{V}}$ &  Family \\
\tableline
  00004 & 2.3615 & 0.0988 & 6.3903 & 468.30 & 26.70 & 0.423 & 0.053 & 00004\\
  00063 & 2.3952 & 0.1206 & 6.2173 & 109.51 &  2.25 & 0.142 & 0.022 & 00004\\
  01273 & 2.3938 & 0.1226 & 6.2289 &   6.77 &  0.15 & 0.299 & 0.035 & 00004\\
  01906 & 2.3736 & 0.0994 & 6.4076 &   8.06 &  0.08 & 0.228 & 0.047 & 00004\\
  01929 & 2.3627 & 0.1141 & 7.0768 &   7.24 &  0.24 & 0.389 & 0.081 & 00004\\
  01933 & 2.3530 & 0.0940 & 6.8229 &   5.48 &  0.07 & 0.454 & 0.044 & 00004\\
  01959 & 2.3161 & 0.0945 & 6.8517 &   7.31 &  0.14 & 0.230 & 0.055 & 00004\\
  01979 & 2.3740 & 0.1015 & 6.5229 &   4.52 &  0.16 & 0.357 & 0.030 & 00004\\
  02011 & 2.3870 & 0.1113 & 6.3730 &   5.19 &  0.65 & 0.463 & 0.100 & 00004\\
  02024 & 2.3254 & 0.0948 & 6.5575 &   8.64 &  0.22 & 0.180 & 0.016 & 00004\\
\hline
\end{tabular}
\label{tab.list}
\end{center}
\end{table}

Most of the families we identify here are analogous to those given by
\citet{nesvornyPDS}, though we identify $28$ new families above our
significance threshold that were not previously known.  In addition,
$24$ families from \citet{nesvornyPDS} are lost in our analysis:
fifteen fell below our cutoff size while nine no longer link at the
QRL level when the albedo components are considered separately.  We
have cross-referenced the lists of family members for overlapping
families in the high-confusion IMB region and removed duplicate
objects in the larger family if they appear in the membership list of
a smaller family to ensure that the small family does not drop below
our cutoff.  This resulted in objects being rejected from the Flora
family list that were linked to Baptistina, and objects being rejected
from the Hertha family list that appeared in the Polana list.

\begin{figure}[ht]
\begin{center}
\includegraphics[scale=0.5]{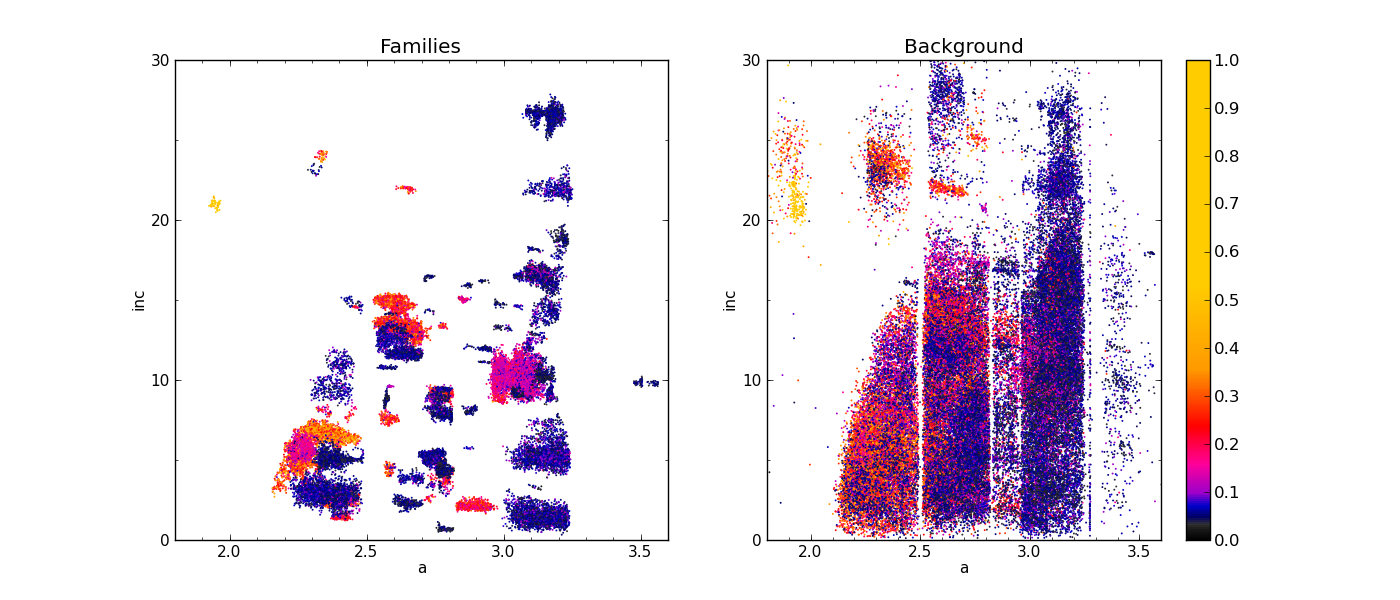}
\caption{Proper inclination (inc, in degrees) vs. proper semimajor axis (a, in AU) for all identified family members (left) and all non-family background objects (right).  The color of the points indicates the albedo of the asteroid as in Figure~\ref{fig.albD} and shown by the color bar.}
\label{fig.aip}
\end{center}
\end{figure}

\begin{figure}[ht]
\begin{center}
\includegraphics[scale=0.5]{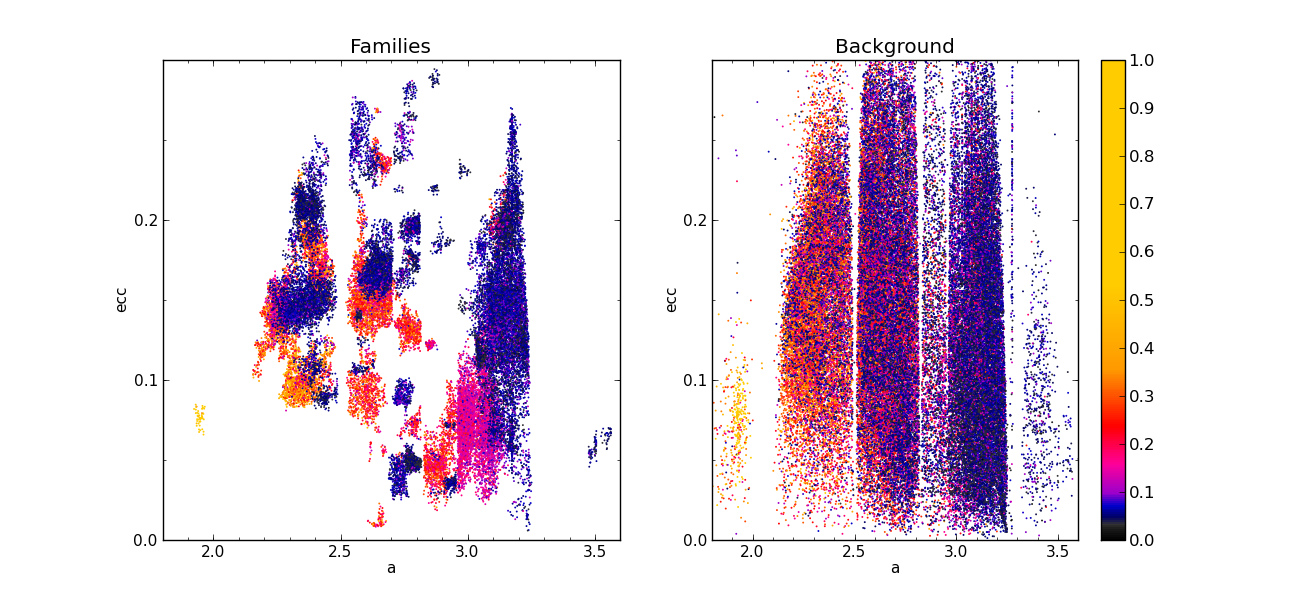}
\caption{Proper eccentricity (ecc) vs. proper semimajor axis (a, in AU) for all identified family members (left) and all non-family background objects (right).  The color of the points indicates the albedo of the asteroid as in Figure~\ref{fig.albD} and shown by the color bar.}
\label{fig.aep}
\end{center}
\end{figure}

\begin{figure}[ht]
\begin{center}
\includegraphics[scale=0.55]{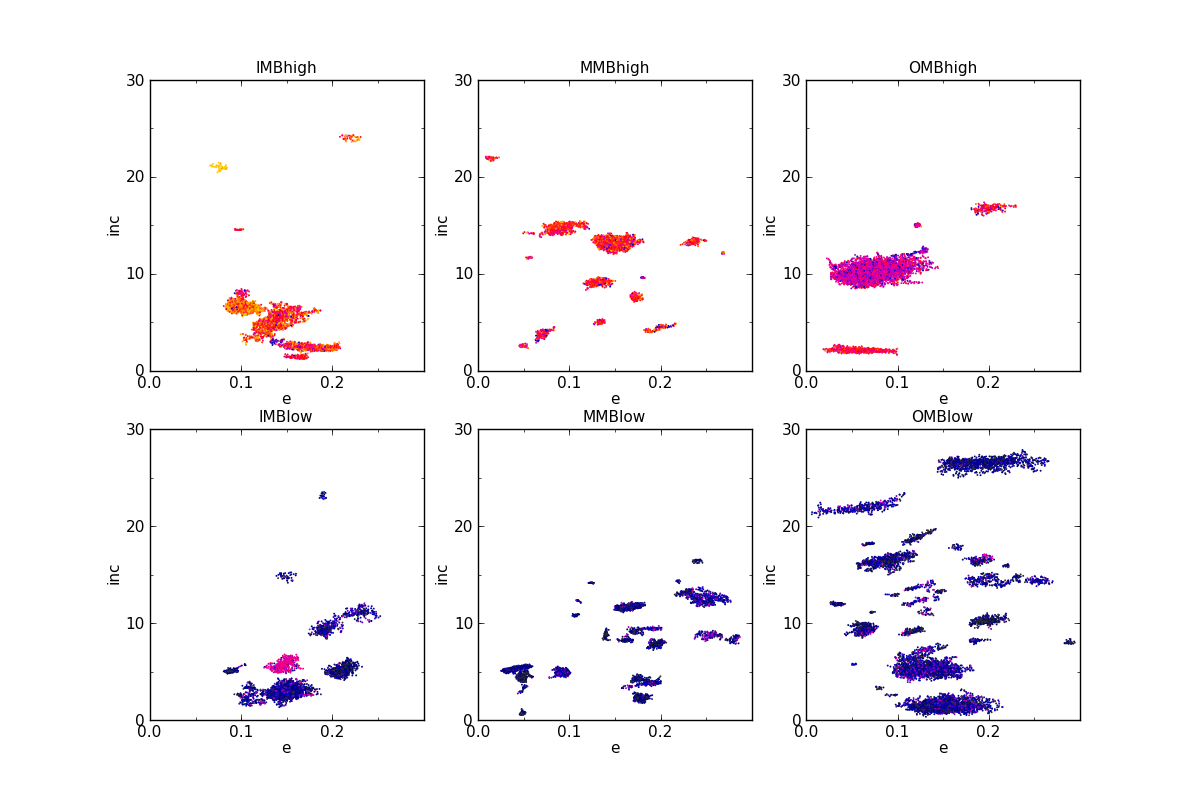}
\caption{Proper inclination (inc, in degrees) vs proper eccentricity
  (ecc) for identified family members for each of the six semimajor
  axis and albedo regions considered in this work (see
  Table~\ref{tab.regions} for the definition of each region).  Color
  indicates the albedo of the asteroid as in Figure~\ref{fig.albD}.}
\label{fig.eiFam}
\end{center}
\end{figure}

\begin{figure}[ht]
\begin{center}
\includegraphics[scale=0.55]{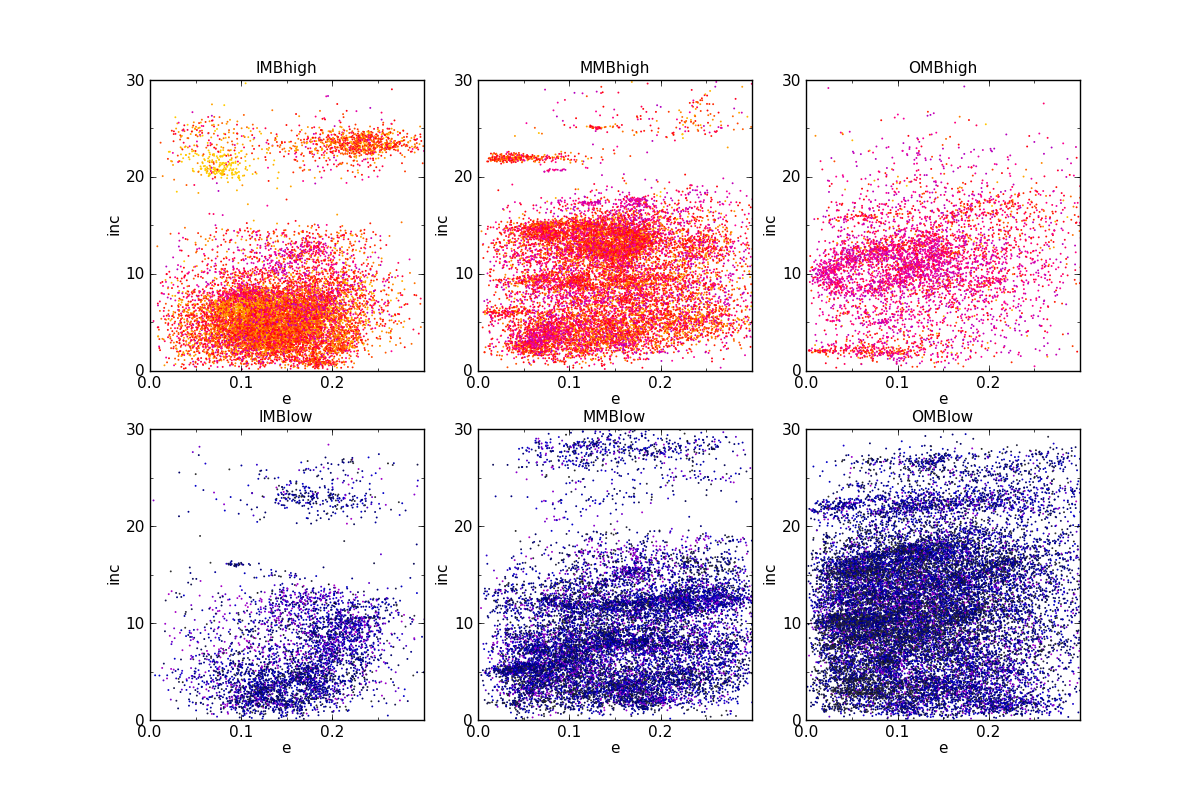}
\caption{The same as Figure~\ref{fig.eiFam} but for background objects
  not part of any identified family.}
\label{fig.eiBack}
\end{center}
\end{figure}

Figures~\ref{fig.aip} and \ref{fig.aep} compare the orbital elements
and albedos of objects identified as part of an asteroid family in
this work to those parameters of the objects that were not linked to
any family and thus assumed to be part of the background.  In
Figure~\ref{fig.eiFam} we separate each semimajor axis and albedo
region and plot the eccentricities and inclinations of the identified
families, while Figure~\ref{fig.eiBack} shows all background objects
in the same fashion.  We see in these plots evidence of halos in the
background distributions around the locations of known large families,
indicating an incomplete identification (and thus removal) of these
families.  For example, the background objects from the OMB$_{high}$
region show a halo near $0<ecc<0.1$ and $8^\circ<inc<13^\circ$
corresponding to unlinked Eos family members.  These halos likely
represent family members that could not be linked to the core of the
family at the QRL for that region, possibly because they have diffused
into the background population due to high initial ejection
velocities, Yarkovsky drift, and/or gravitational perturbation.  We
note two features in the IMB: a number of low albedo objects from the
Polana family have been dragged into the IMB$_{high}$ by the Hertha
family from the region of albedo overlap included in the high albedo
cutoff; and the moderately-high albedo Baptistina family is plotted in
the IMB$_{low}$ region as this family blends with Flora and cannot be
distinguished in the high albedo population.

As is clearly shown in these plots, careful fine-tuning of the cutoff
velocity and extraction of each individual family could increase the
population of known families significantly.  This is particularly true
for the high inclination regions where spatial densities are low;
using a single QRL for the high and low inclination components of each
region is likely resulting in a failure to identify a large number of
family members at high inclinations (e.g. near the Hungaria family at
$a\sim1.8~$AU, $inc\sim20^\circ$, and $ecc\sim0.08$).  However,
subdividing each region's QRL and fine-tuning each family's cutoff
velocity increases the subjectivity of the family determination;
future work will attempt to investigate this more quantitatively.
It is likely that this extension to the present work will increase the
fraction of the MBAs that are members of a family significantly,
possibly to $\sim50\%$ or more which would mean that most asteroids
are by-products of a catastrophic collision or large cratering event
\citep[cf.][]{bottke05}.

\clearpage

\citet{gilhutton06} identified 13 nominal families at high
inclinations using proper orbital elements and HCM.  Seven of those
correspond to families that we identify here, one was linked by our
routine but was too small to pass the reliability cuts.  The remaining
five found by \citet{gilhutton06} were not linked in our method,
however four of those had only a small number of members.  In
particular, we can not link any of the three families those authors
identify in the IMB below our nominal QRL.  We are also unable to
identify any of the new families presented in \citep{novakovic11}.
These differences in family lists are a result of our application of a
single QRL to both high and low inclination IMB objects despite the
large differences in spatial density between the two populations.  We
anticipate that with further refinement of the QRL for high
inclination objects these families will be recoverable.

Recently, \citet{broz13} published a revised list of asteroid family
members, rerunning the HCM linking routine over the latest list of
asteroids.  The majority of the families those authors present are
similar to the ones we discuss here, and most of their families we do
not find are because they are too small, at high inclination, or do
not link at our QRL when confusion with background sources is reduced.
There are two specific examples of differences between this work and
theirs that we comment on here.  First, \citet{broz13} identify a
family around (1044) Teutonia with $1950$ members, however in our data
this family does not link significantly at a level below our QRL, and
at the QRL only links to $\sim65$ objects.  Second, they identify a
family around (2085) Henan with $946$ members, however as with
Teutonia this family does not link below our QRL level.  We interpret
both of these cases as instances where confusion with background
sources is eliminated when albedos are separated and thus linkages no
longer become significant.

\begin{figure}[ht]
\begin{center}
\includegraphics[scale=0.6]{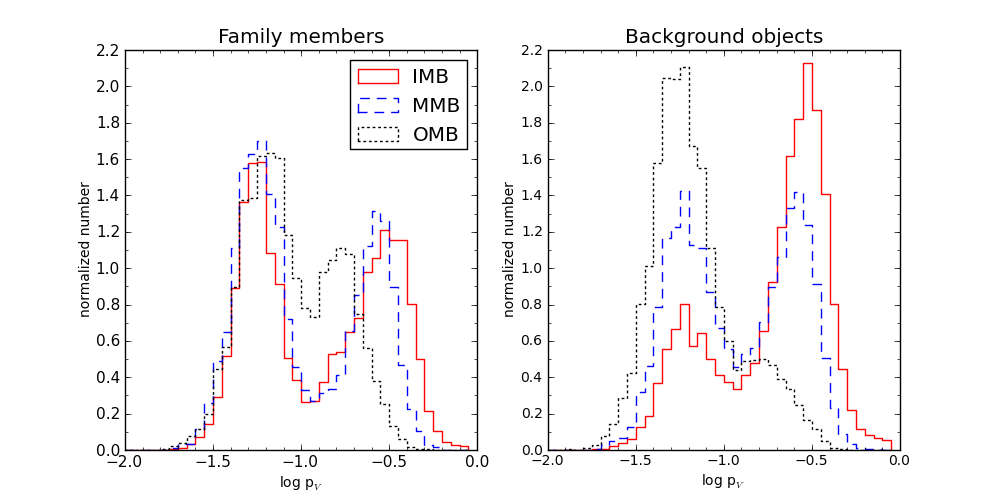}
\caption{Albedo distribution for the IMB (red solid), MMB (blue
  dashed), and OMB (black dotted) for all identified family members
  (left) and all background objects (right)}
\label{fig.albHist}
\end{center}
\end{figure}

We compare in Figure~\ref{fig.albHist} the albedo distribution of
asteroids linked to a dynamical family to the distribution for
background objects for each of the three orbital regions considered.
While the family members show roughly equal contributions from high
and low albedo objects in all three regions, the background objects
show the opposite trend.  The IMB background is dominated by the
higher albedo component, while the OMB background has only a minimal
contribution from these objects.  This division is likely to become
even more extreme once the halo objects are properly accounted for, as
these make up a large fraction of the IMB$_{low}$ and OMB$_{high}$
objects remaining in the background population.

\begin{figure}[ht]
\begin{center}
\includegraphics[scale=0.6]{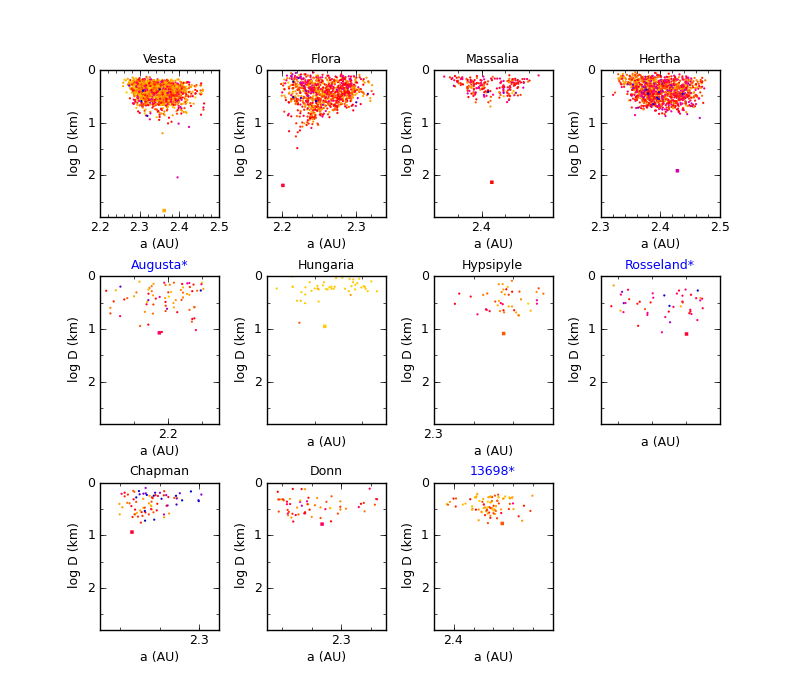}
\caption{Diameter vs semimajor axis petal plots for all significant
  families in the high albedo inner Main Belt region.  The parent body
  is indicated as the larger square point.  Names with a `*' suffix
  have ambiguous parent bodies.  Colors of the points indicate the
  albedo of the asteroid as in Figure~\ref{fig.albD}.}
\label{fig.petalIH}
\end{center}
\end{figure}

\begin{figure}[ht]
\begin{center}
\includegraphics[scale=0.6]{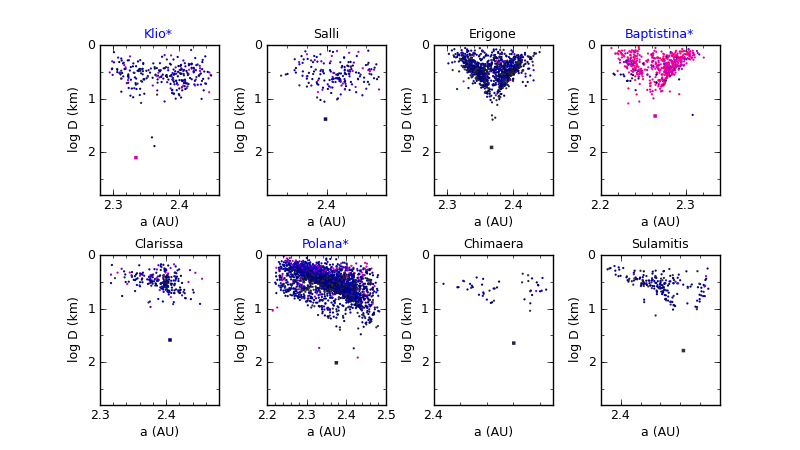}
\caption{The same as Figure~\ref{fig.petalIH} but for the low albedo
  inner Main Belt.}
\label{fig.petalIL}
\end{center}
\end{figure}

\begin{figure}[ht]
\begin{center}
\includegraphics[scale=0.6]{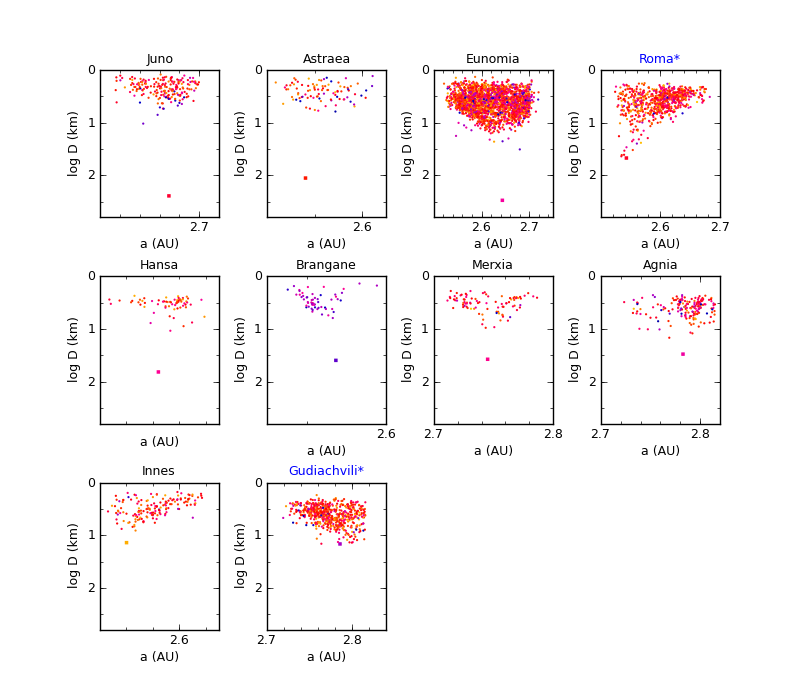}
\caption{The same as Figure~\ref{fig.petalIH} but for the high albedo
  middle Main Belt.}
\label{fig.petalMH}
\end{center}
\end{figure}

\begin{figure}[ht]
\begin{center}
\includegraphics[scale=0.6]{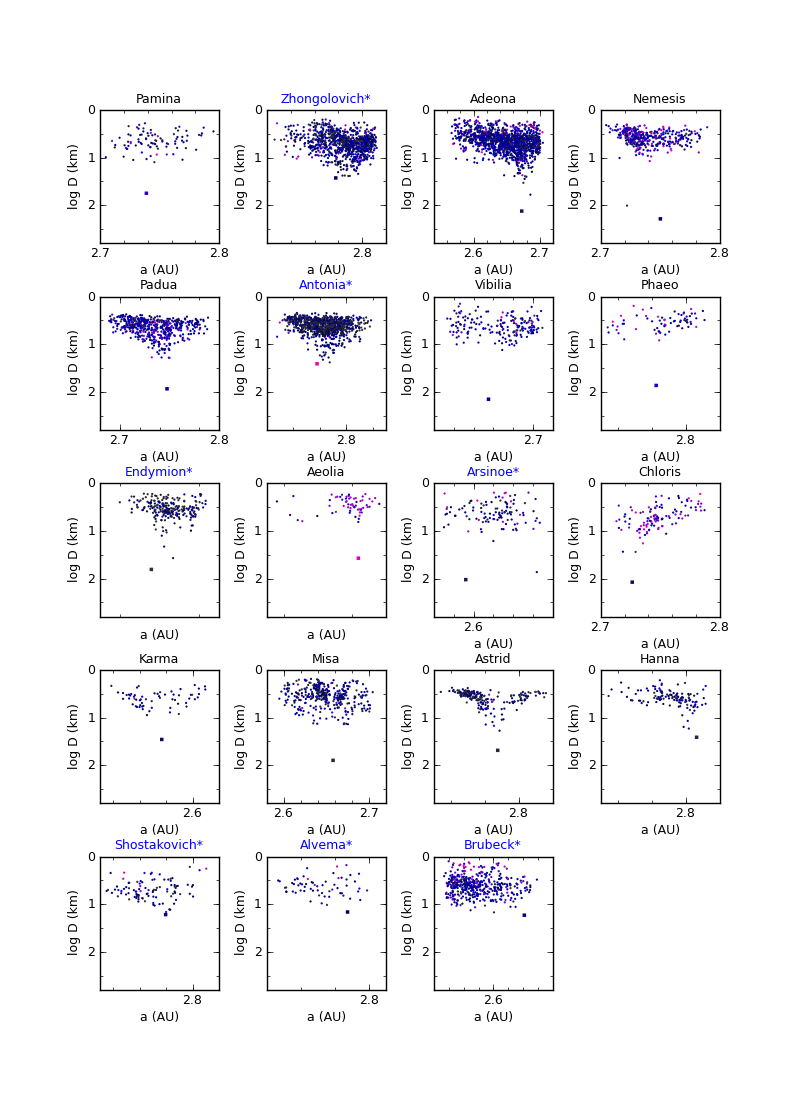}
\caption{The same as Figure~\ref{fig.petalIH} but for the low albedo
  middle Main Belt.}
\label{fig.petalML}
\end{center}
\end{figure}

\begin{figure}[ht]
\begin{center}
\includegraphics[scale=0.6]{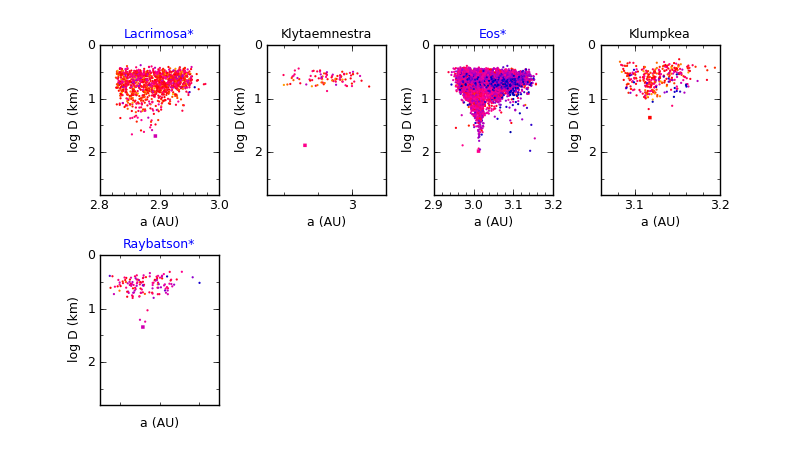}
\caption{The same as Figure~\ref{fig.petalIH} but for the high albedo
  outer Main Belt.}
\label{fig.petalOH}
\end{center}
\end{figure}

\begin{figure}[ht]
\begin{center}
\includegraphics[scale=0.6]{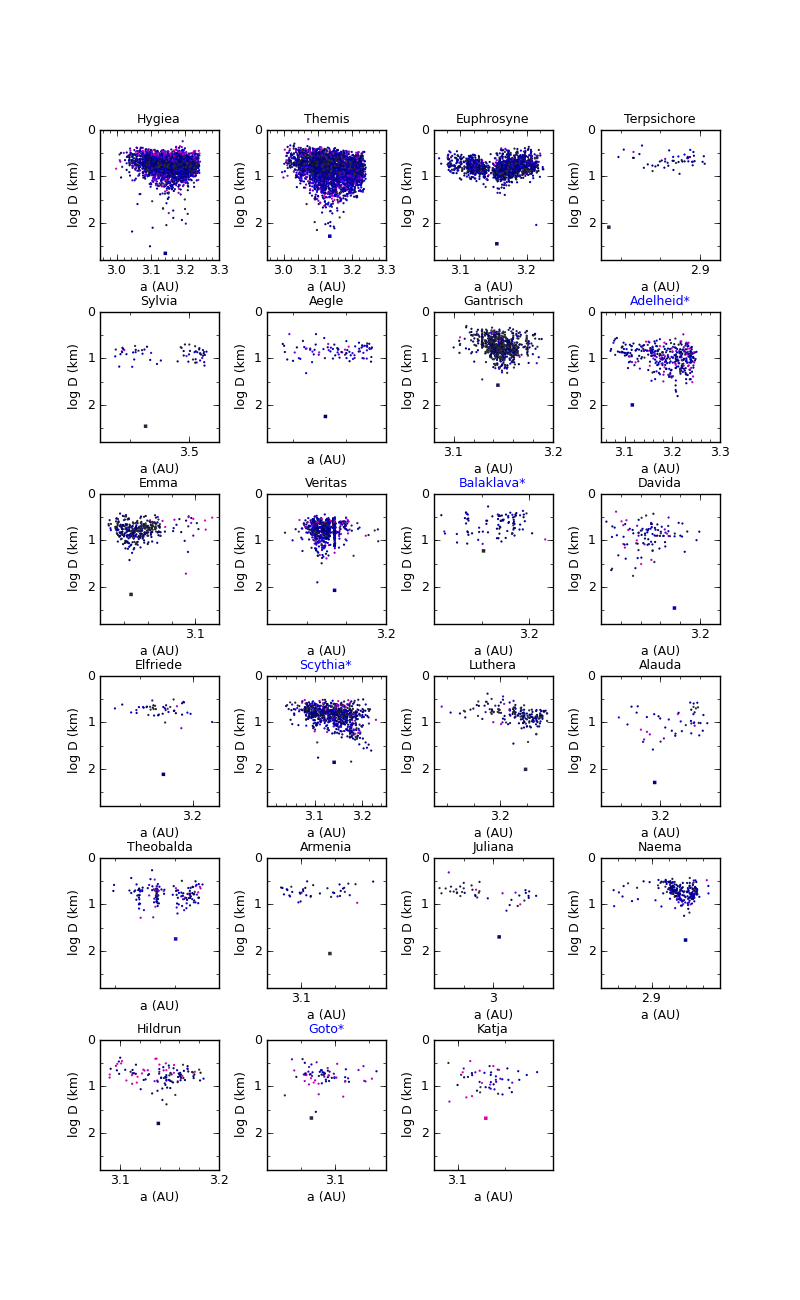}
\caption{The same as Figure~\ref{fig.petalIH} but for the low albedo
  outer Main Belt.}
\label{fig.petalOL}
\end{center}
\end{figure}

In Figures~\ref{fig.petalIH}-\ref{fig.petalOL} we show the individual
petal plots for each of the families we identify, separated into
regions.  Each subplot is labeled with the name of the largest body;
names in blue font with a `*' suffix indicate cases of ambiguous
parent bodies, either because there is no clear largest body
(e.g. Endymion) or there is a small group of objects of similar sizes
that could be the parent or represent a completely shattered parent
(e.g. the Eos family, where (221) Eos, (639) Latona, and (579) Sidonia
are clustered at similar semimajor axes with diameters of D$=96~$km,
$89~$km, and $86~$km respectively).  In some cases it is clear that
the listed family likely represents two overlapping families
(e.g. Emma).  We also indicate Baptistina as an ambiguous case given
the results found in \citet{masiero12bap}.  

We note that as we have chosen to identify families by the name of the
largest member, rather than the lowest numbered object, in some cases
well-known families have changed name.  A prime example of this is the
Lacrimosa family, which encompasses the well-known Koronis family.  As
(208) Lacrimosa has a larger diameter than (158) Koronis (D$=50~$km vs
D$=39~$km) our naming system assigns the family to Lacrimosa, but this
case is another example of a family with an ambiguous parent.

\clearpage

\clearpage

\subsection{Family Physical Properties}
\label{sec.phys}

Using the high-confidence family lists we investigate the physical
parameters of each family including the characteristic size, raw
size-frequency distribution (SFD) and mean albedo as they relate to
number of family members identified and size of the largest remnant.
We note that as shown in Figure~\ref{fig.albD} our sample, limited by
the biases imposed predominantly by visual band detection, is not
complete to the same sizes for high- and low-albedo objects.  This
means that the minimum and median sizes of low albedo families will be
shifted to larger values than for high albedo families.  Additionally,
for a given family the wing of the albedo distribution at higher
values will be better sampled than the wing at lower values.  Also
note that our imposed division of albedos into two components will
remove objects that have albedos very different from the bulk of the
family due to incorrect measurements or large physical differences.
These biases will affect the measurement of SFD as well, so care in
interpretation is required.

\begin{figure}[ht]
\begin{center}
\includegraphics[scale=0.6]{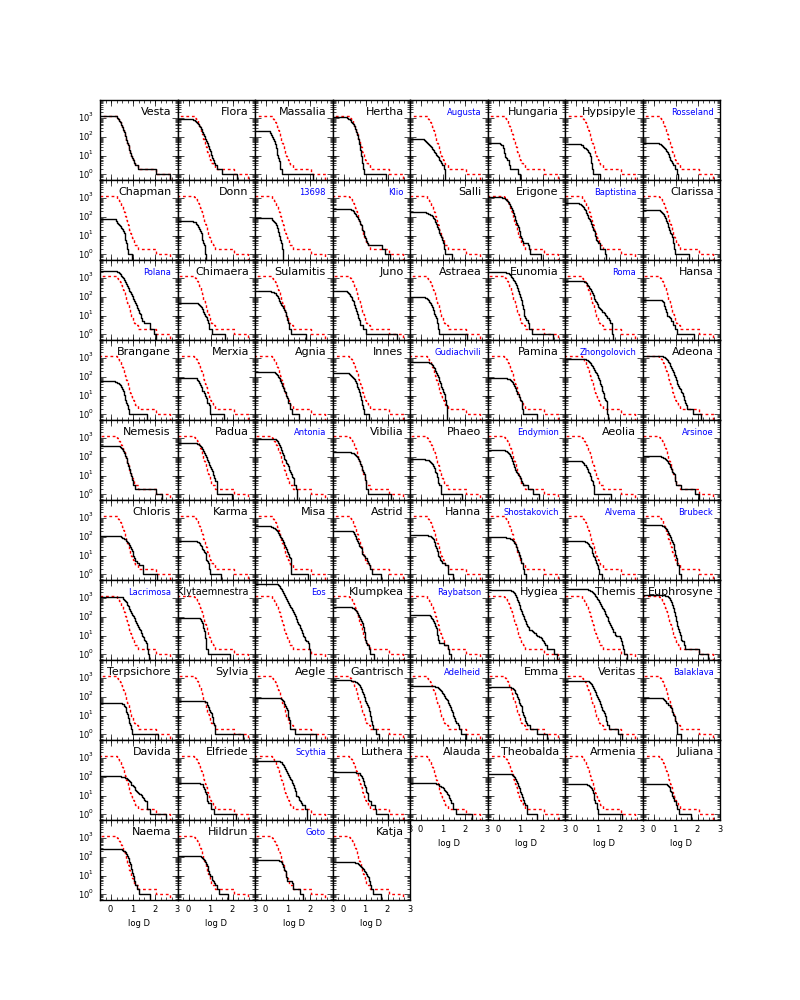}
\caption{Cumulative raw size distribution of family members identified
  in this study for the $76$ families presented in
  Table~\ref{tab.stats}.  The red dashed line shows the distribution
  for the Vesta family, for comparison.  Family names in blue indicate
  cases where the parent body is uncertain.  As discussed in the text,
  selection biases from visible light surveys have not been quantified
  which will affect interpretation of these distributions.}
\label{fig.dDist}
\end{center}
\end{figure}

\begin{figure}[ht]
\begin{center}
\includegraphics[scale=0.6]{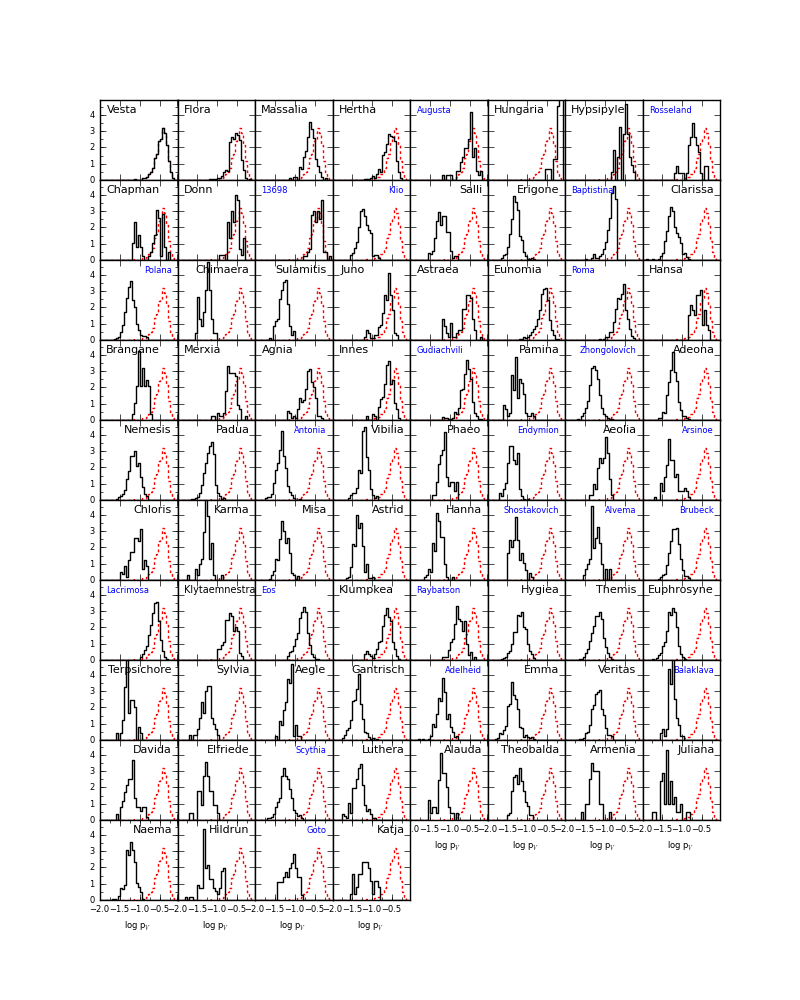}
\caption{Differential raw distribution of albedos for the $76$
  families presented in Table~\ref{tab.stats}.  The red dashed line
  shows the distribution for the Vesta family, for comparison.  Family
  names in blue indicate cases where the parent body is uncertain.  As
  discussed in the text, selection biases from visible light surveys
  have not been quantified which will affect interpretation of these
  distributions.}
\label{fig.albDist}
\end{center}
\end{figure}

We present the observed cumulative size distribution for each family in
Figure~\ref{fig.dDist}, and the differential albedo distribution for
each family in Figure~\ref{fig.albDist}.  Approximately two-thirds of
the families show a single parent body much larger than the remaining
members, while the remaining families do not show a dominant remnant.
This may be indicative of differences between cratering events and
super-catastrophic disruptions as discussed by \citet{durda07},
however in the latter families we cannot rule out cases where the
largest remnant is present but did not link to the family due to the
various cutoffs imposed.  As opposed to what was seen in Mas11, where
$\sim25\%$ of the previously published families had non-uniform albedo
distributions, almost none of the families presented here show any
bimodality in their albedo distributions.  The few exceptions
(e.g. Chapman, Astraea) are expected to be artifacts that will be
removed with future planned revisions of the family selection cutoffs.
Given our separation of the regions into high- and low-albedo
components, this result is not unexpected.

Using these distributions, we can determine initial physical
properties for each family with the caveat that selection biases
imposed by visible light surveys have not yet been removed and so will
bias our results as well.  We fit a Gaussian profile to the albedo
distribution of each family to derive the mean albedo ($p_V$) and
Gaussian width ($\sigma_{p_V}$), which is $20\% - 30\%$ of the mean
for nearly all of the families we observe.  This width is comparable
to the uncertainty we expect in the albedo determination from the
NEOWISE data \citep[Mas11,][]{mainzer11cal}.  Using the diameter
distribution, we can size-sort the linked family to find the diameter
of the largest body ($D_{max}$), the median diameter of the observed
sample ($D_{med}$), as well an initial estimate of the observed, raw SFD.

\clearpage

To determine the observed SFD we fit a function of $N = D^{\alpha}$ to
each cumulative distribution, using only those bins with more than
five objects (to avoid errors induced by a few large fragments in a
cratering scenario) and less than half of the sample (to minimize
errors due to the incomplete catalogs).  Using a least-squares
minimizer we derive the best-fit slope parameter ($\alpha_{SFD}$) and
the $1\sigma$ error on that slope ($\sigma_\alpha$).  We see in our
sample a range of slope parameters, from $\alpha\sim-1.5$ at the
shallowest to $\alpha\sim-5$ at the steepest.  SFD slope can be used
to study the conditions of the breakup event \citep[e.g. the curve of
  the SFD can trace cratering vs. catastrophic
  disruption,][]{durda07}, however these slopes can be modified over
time as collisional grinding alters the initial post-breakup
population, flattening the SFD over time \citep{marzari95}.
Additionally, if the parent body prior to impact had a shattered
interior structure, this will also change the expected SFD of the
family by increasing the frequency of reaccumulation of large family
members \citep{michel04}.

The specific bins chosen for fitting will alter the fitted slope
depending on the shape of the SFD.  Tests on our data show that
varying the lower limit over a range from five to fifty objects
typically causes the fitted slopes to become shallower by
$\Delta\alpha\sim0.1-0.3$, with the effect becoming more pronounced
for smaller families as the sample size decreases.  This indicates
that most families have steeper size distributions in the biggest
objects than in the majority of the population, however this is
probably at least partially a result of survey incompleteness at
smaller diameters.  Conversely, when we test the effect of changing
the upper limit for fitting from half the population down to only the
largest $25\%$, we see that the fitted slopes generally move away from
a value of $\alpha\sim-2.5$ as fewer objects are included in the fit
\citep[cf.][]{tanga99}.  This shift is typically $\Delta\alpha<0.5$
however for the most extreme case (the Hertha family) it is as large
as $\Delta\alpha=1.3$.  As most families have measured slopes of
$\alpha<-2.5$ this means that their SFDs become steeper as objects are
removed.  This effect may also be a result of survey incompleteness at
the smallest sizes, or may be tracing either a collisional equilibrium
at smaller sizes with the background population which has
$\alpha\sim-2.5$ (Mas11) or an increasing fraction of non-family
members being included in family lists at these sizes.

This uncertainty in the fit to the observed SFD is compounded by the
visible-light survey biases that are incorporated into the optical
measurements we use to determine albedo.  Objects without albedos do
not pass our initial cuts for inclusion into our data set, thus our
sample will include some effect from these biases, which will be
particularly pronounced for smaller, lower albedo objects.  These
biases will result in systematic and potentially large errors in the
fits to the observed SFD.  Debiasing of the MBA background and family
populations is critical to measuring the true SFDs and physical
parameters for asteroid families, and as such values given here should
be regarded as preliminary.

We present in Table~\ref{tab.stats} the determined albedo, diameter,
and slope parameters for each of the $76$ significant families we
identify.  These values are subject to large biases due to incomplete
sampling of various subpopulations within the MBAs, so care in
interpretation is required.  We note that as families indicated by a
`*' suffix on their name have ambiguous parent bodies, the size of the
largest linked object is not necessarily indicative of the size of the
parent or largest remnant.  

\begin{figure}[ht]
\begin{center}
\includegraphics[scale=0.5]{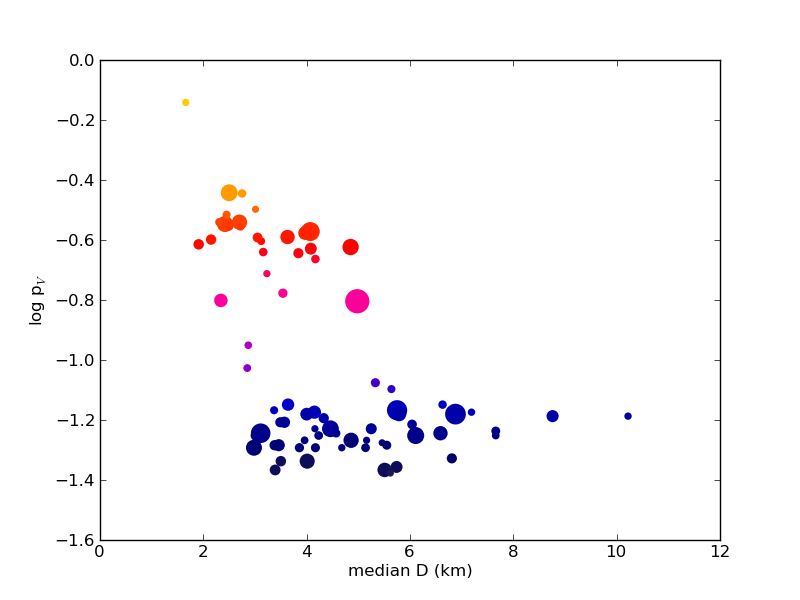}
\caption{Characteristic albedo for each family compared to the median
  diameter of all objects linked to that family.  The size of the
  points indicates the number of objects in the family, while the
  color also traces albedo, following Figure~\ref{fig.albD}.  As
  expected, higher albedo families have small median diameters as
  selection effects against smaller, lower albedo asteroids imposed by
  visible light surveys have not been removed.}
\label{fig.compAlbD}
\end{center}
\end{figure}

We show in Figure~\ref{fig.compAlbD} a comparison of the mean family
albedo to the median diameter.  As expected from the biases against
smaller, lower albedo objects imposed by visible light surveys used to
create the catalog our sample is drawn from, high albedo families have
a smaller median diameter than low albedo families.  Despite this
bias, nearly two-thirds of the observed families are low albedo.  As
catalogs of asteroids become more complete at small sizes for low
albedo objects we expect that the number of low albedo families
identified will increase further.  We also see indications for four
distinct albedo classes in the families: low albedo ($p_V\sim0.06$),
moderate albedo ($p_V\sim0.16$), high albedo ($p_V\sim0.25$) and the
very high albedo Hungaria family ($p_V>0.40$).  We note that while
updates to the measured magnitudes of the Hungaria family including
better fits of $H$ and $G$ are expected to reduce the $H$ magnitudes
and thus the determined albedos (B. Warner, 2012, private
communication), these family members still typically have $p_V>0.40$
and are expected to remain narrowly distributed.

\begin{figure}[ht]
\begin{center}
\includegraphics[scale=0.5]{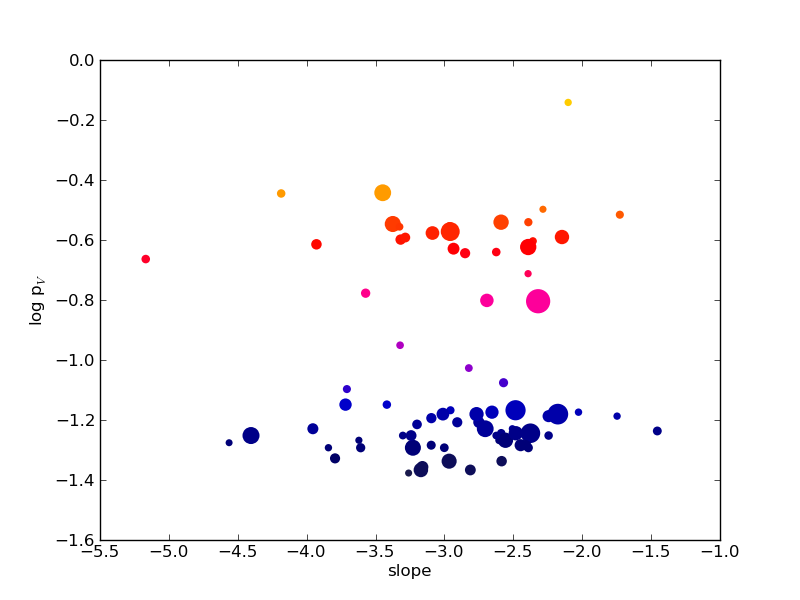}
\caption{The same as Figure~\ref{fig.compAlbD}, now showing albedo
  vs. SFD slope parameter.  For the largest families the error on the
  slope fit is smaller than the size of the point, however systematic
  errors due to survey bias have not been included.}
\label{fig.compSlopeAlb}
\end{center}
\end{figure}

If we take albedo as a coarse tracer of the parent body's composition
\citep[cf.][]{mainzer11tax} we can search for any influence this has
on the resultant breakup.  We show in Figure~\ref{fig.compSlopeAlb} a
comparison between the characteristic family albedo and the fitted
observed SFD slope.  If different compositions traced different
cohesion strengths, we would expect to see a correlation between these
two parameters, however we see no such relationship.  This may
indicate that albedo does not correlate with internal structure, or
that impact geometry and/or energy play the dominant role in shaping
the SFD of the family, as shown by \citet{durda07}, however debiasing
must be performed before definite conclusions can be drawn.

\begin{figure}[ht]
\begin{center}
\includegraphics[scale=0.5]{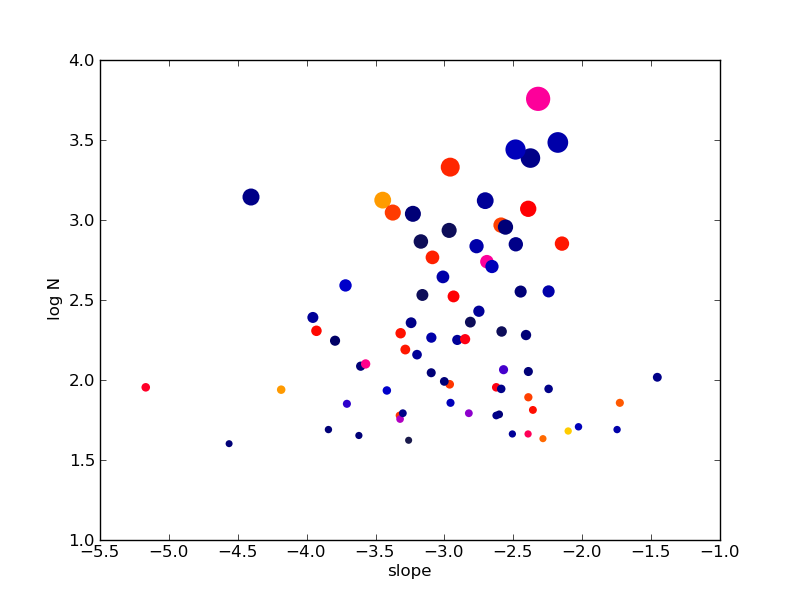}
\caption{The same as Figure~\ref{fig.compAlbD}, now showing size of
  the linked family vs. observed raw SFD slope parameter.  For the
  largest families the error on the slope is smaller than the size of
  the point, however systematic errors due to survey bias have not
  been included.}
\label{fig.compSlopeN}
\end{center}
\end{figure}

Figure~\ref{fig.compSlopeN} shows the comparison between the observed
raw SFD slope and the size of the linked family.  Clearly, families
with smaller numbers of objects will have more uncertain SFD slopes.
However, there does appear to be a trend toward shallower slopes for
larger families.  We note that in a diameter-limited survey such as
NEOWISE fewer objects are expected to link to families with steep SFDs
than shallow SFDs, as a larger fraction of the family members will be
at or below the survey detection threshold.  This may be a
contributing factor to the lack of large, steep SFD families.
Conversely, biases due to variable survey completeness can induce
systematic errors.  As steep families are thought to be young, and a
large family indicates an energetic collision, these families are
expected to be rare, confusing any interpretation of their absence.  A
survey probing a smaller size range would preferentially fill in the
membership lists of steep SFD families allowing better measurements of
their SFDs to be made.

\citet{tanga99} and \citet{durda07} showed that steeper slopes tend to
indicate families that originated from a cratering event, while
shallower slopes tend to follow disruptive impacts.  This suggests
that in the diameter range we are sensitive to, most of the
numerically largest families in the Main Belt formed from disruptive
events.  An exception to this trend is the (31) Euphrosyne family,
which has $\sim1400$ members but one of the steepest slopes measured
with $\alpha_{SFD}=-4.4$.  This unusually steep observed SFD may be
indicative of a glancing impact between two large bodies resulting in
a large cratering event.  We note that this observed SFD is
significantly steeper than what is observed for (4) Vesta
($\alpha=-3.45$, the next steepest SFD in this family size range),
which is known to have undergone two massive cratering events from the
results of the Dawn mission \citep{russell12}.  Conversely, the
observed SFD of the Euphrosyne family may trace a low-speed collision
where reaccumulation onto the parent was highly efficient.  In this
scenario only those objects with the highest ejection velocities,
preferentially the smallest ejecta, would fail to reaccrete and would
become independent family members.  As grazing impacts have a much
lower frequency of binary production than slower, head-on impacts
\citep{durda04}, a search for binaries in the Euphrosyne family could
differentiate between these two scenarios.  However, until proper
debiasing of the MBA population is complete and debiased SFDs can be
measured, these possible formation scenarios are speculative.

\section{Conclusions}

In order to identify new asteroid families and improve the lists of
previously known families, we have used the databases of asteroid
diameters and albedos provided by the WISE/NEOWISE and IRAS surveys as a
method of incorporating asteroid physical properties into family
determination.  After splitting the MBAs into two separate but
overlapping albedo populations and three orbital regions, we have
applied the Hierarchical Clustering Method to these subgroups.  This
technique is particularly useful for rejecting interlopers from family
lists, including objects that would otherwise be misidentified as the
family's largest remnant.  We identify $76$ families with more than
$40$ members in our high-confidence set, of which $28$ were previously
unreported.  One third of these families do not have an obvious parent
asteroid that links to the family. Approximately $35\%$ of MBAs link
into a family, however there are clear limitations to our method
apparent at high inclinations and in crowded regions.  Carefully tuned
family extraction will likely increase this percentage significantly.

The observed albedo distribution of the $38,298$ identified family
members differs significantly from the distribution of the background
asteroids, where the background shows a more extreme change from the
inner- to outer-Main Belt.  We see no correlation between the slope of
the observed, raw size frequency distribution of a family and its mean
albedo.  However, we note that a full accounting for the selection
biases imposed by the visible light surveys that discovered the
objects has not yet been performed, and an unquantified number of
objects are thus missing from each family; these biases must be
accounted for before firm conclusions about the SFDs for each family
can be made.  Of the observed raw SFDs, we note a trend that families
with larger numbers of objects tend to have more shallow slopes.  One
possible explanation is that these families were created could by
catastrophic disruption or are very old and have been ground down by
collisional processing over very long timescales.  An exception to
this trend is the (31) Euphrosyne family which has one of the steepest
size frequency distributions measured, but is also one of the largest
families observed.  This could be the result of a cratering impact
between two large bodies, a recent giant impact, or both.  While this
work only includes approximately one quarter of the known MBA
population (those observed by NEOWISE and IRAS), it sets the stage for
improving our understanding of the creation and evolution of MBA
families.  Future work will improve our family identification routines
to better differentiate family members from background objects and
debias the input catalog, improving our family lists and setting the
stage for the determination of ages for the majority of families in
the Main Belt.

\section*{Acknowledgments}
JM was partially supported by an appointment to the NASA Postdoctoral
Program at JPL, administered by Oak Ridge Associated Universities
through a contract with NASA.  The authors thank the referee Alberto
Cellino for comments that greatly improved the paper, and John
Dailey for helpful discussions on implementation of algorithms.  This
publication makes use of data products from the Wide-field Infrared
Survey Explorer, which is a joint project of the University of
California, Los Angeles, and the Jet Propulsion Laboratory/California
Institute of Technology, funded by the National Aeronautics and Space
Administration.  This publication also makes use of data products from
NEOWISE, which is a project of the Jet Propulsion
Laboratory/California Institute of Technology, funded by the Planetary
Science Division of the National Aeronautics and Space Administration.
This research has made use of the NASA/IPAC Infrared Science Archive,
which is operated by the Jet Propulsion Laboratory, California
Institute of Technology, under contract with the National Aeronautics
and Space Administration.  We gratefully acknowledge the extraordinary
services specific to NEOWISE contributed by the International
Astronomical Union's Minor Planet Center, operated by the
Harvard-Smithsonian Center for Astrophysics, and the Central Bureau
for Astronomical Telegrams, operated by Harvard University.

\end{document}